\documentclass[reprint,superscriptaddress,preprintnumbers,amsmath,amssymb,aps,prd,tightenlines,longbibliography,nofootinbib]{revtex4-2}

\usepackage{graphicx}
\usepackage{xcolor}
\usepackage[sort&compress]{natbib}
\usepackage{amsmath,amssymb,bm,bbm,slashed,subdepth,amsfonts}
\usepackage{xr-hyper}
\usepackage[colorlinks=true
,urlcolor=blue
,anchorcolor=blue
,citecolor=blue
,filecolor=blue
,linkcolor=red
,menucolor=blue
,linktocpage=true
,pdfproducer=medialab
,pdfa=true
]{hyperref}
\usepackage{cleveref}
\usepackage{enumerate}
\usepackage{epsfig, subfigure}
\usepackage{setspace}
\usepackage{booktabs, tabularx}
\usepackage{units}
\usepackage{placeins}
\usepackage{multirow}
\usepackage{mathtools}
\usepackage[normalem]{ulem}

\newcommand{\be}{\begin{equation}}
\newcommand{\ee}{\end{equation}}
\newcommand{\bea}{\begin{equation}\begin{aligned}}
\newcommand{\eea}{\end{aligned}\end{equation}}

\newcommand{\la}{\langle}
\newcommand{\ra}{\rangle}
\newcommand{\vt}{\vert}

\newcommand{\bx}{\mathbf{x}}
\newcommand{\bd}{\mathbf{d}}
\newcommand{\bk}{\mathbf{k}}
\newcommand{\bq}{\mathbf{q}}
\newcommand{\bv}{\mathbf{v}}
\newcommand{\bn}{\mathbf{n}}
\newcommand{\bzeta}{\boldsymbol \zeta}

\newcommand{\DM}{{\scriptscriptstyle \textrm{DM}}}
\newcommand{\eV}{\textrm{eV}}

\begin{document}

\title{A Quantum Description of Wave Dark Matter}
 
% =============================================================================

\author{Dhong Yeon Cheong}
\email{dhongyeon@yonsei.ac.kr}
\affiliation{Department of Physics and IPAP, Yonsei University, Seoul 03722, Republic of Korea}

\author{Nicholas L. Rodd}
\email{nrodd@lbl.gov}
\affiliation{Theory Group, Lawrence Berkeley National Laboratory, Berkeley, CA 94720, USA}
\affiliation{Berkeley Center for Theoretical Physics, University of California, Berkeley, CA 94720, USA}

\author{Lian-Tao Wang}
\email{liantaow@uchicago.edu}
\affiliation{Enrico Fermi Institute, The University of Chicago, 5640 S Ellis Ave, Chicago, IL 60637, USA}
\affiliation{Department of Physics, The University of Chicago, 5640 S Ellis Ave, Chicago, IL 60637, USA}
\affiliation{Kavli Institute for Cosmological Physics, The University of Chicago, 5640 S Ellis Ave, Chicago,
IL 60637, USA}

\begin{abstract}
We outline a fundamentally quantum description of bosonic dark matter (DM) from which the conventional classical-wave picture emerges in the limit $m \ll 10~\eV$.
As appropriate for a quantum system, we start from the density matrix which encodes the full information regarding the possible measurements we could make of DM and their fluctuations.
Following fundamental results in quantum optics, we argue that for DM it is most likely that the density matrix takes the explicitly mixed form of a Gaussian over the basis of coherent states.
Deviations from this would generate non-Gaussian fluctuations in DM observables, allowing a direct probe of the quantum state of DM.
Our quantum optics inspired approach allows us to rigorously define and interpret various quantities that are often only described heuristically, such as the coherence time or length.
The formalism further provides a continuous description of DM through the wave-particle transition, which we exploit to study how density fluctuations over various physical scales evolve between the two limits and to reveal the unique behavior of DM near the boundary of the wave and particle descriptions.
\end{abstract}

\maketitle

%%%%%%%%%%%%%%%%%%%%%%%%%%%%%%%

The vastness of the allowed dark-matter (DM) mass range allows for dramatic variations in its behavior as one moves through the landscape.
The most famous of these is the wave-particle transition, which occurs near the location of the Earth at a DM mass of $m \sim 10\,\eV$.
A heuristic justification for the transition is as follows.
The number density of DM is determined from $n = \rho/m$, whereas a rough measure for the volume occupied by each state is the de Broglie volume, $V_\textrm{dB} \simeq (m v)^{-3}$.
Here $\rho \simeq 0.4\,\textrm{GeV/cm}^3$ and $v \simeq 10^{-3}$ are the local DM energy density and mean speed, respectively.
Accordingly, the DM states overlap when $n V_\textrm{dB} = \rho/m^4 v^3 \gtrsim 1$, or equivalently for $m \lesssim 10\,\eV$.
As states begin to overlap, one argues that a collective description of DM as a classical wave is more appropriate.
Focusing on the case of a scalar field, $\phi$, we describe DM as an oscillating field with a frequency dictated by its mass and an amplitude fixed to match the local density,
\be
\phi(t) \simeq \frac{\sqrt{2\rho}}{m} \cos(mt).
\label{eq:naive}
\ee

The above discussion underpins the search for wave DM candidates, such as the axion~\cite{Preskill:1982cy,Abbott:1982af,Peccei:1977hh,Peccei:1977ur,Weinberg:1977ma,Wilczek:1977pj,Dine:1981rt,Dine:1982ah,Zhitnitsky:1980tq,Kim:1979if,Shifman:1979if}.
It also raises many questions.
How accurate is the result in Eq.~\eqref{eq:naive}?
What are the leading order corrections to this result?
How can it be derived rigorously?
Does the field simply oscillate or can it exhibit large fluctuations around its mean value?
When $m \sim 10\,\eV$ and DM is neither deep in the wave or particle regime, how should it be described?
The goal of the present work is to resolve such questions by providing a fundamentally quantum mechanical description of DM from which the wave picture emerges.\footnote{Detailed studies of wave DM often employ more sophisticated models than presented in Eq.~\eqref{eq:naive}.
For instance, Ref.~\cite{Foster:2017hbq} introduced a model to accommodate the finite dispersion of DM by treating the field as a sum of cosines with frequencies weighted by the underlying energy distribution of DM.
Nevertheless, we emphasize that even such refinement do not answer the fundamental questions raised.}

The main insight used to achieve this goal is to realize that similar questions have already been answered in a different context.
In particular, the field of quantum optics provides a fully quantum mechanical description of electromagnetic radiation.
Many of the results from that field can be extended to any bosonic energy density, and therefore apply to ultralight DM.
Indeed, there have already been broad efforts to exploit quantum optics for various aspects of wave DM; it has been widely used in the work of Kim~\cite{Kim:2021yyo,Kim:2022mdj,Kim:2023pkx,Kim:2023pvt,Kim:2023kyy,Kim:2024xcr} and a number of other groups, see e.g. Refs.~\cite{Derevianko:2016vpm,Masia-Roig:2022net,Blinov:2024jiz,Bernal:2024hcc}.
Our focus here is on systematizing the discussion and outlining both where the foundations for wave DM can be placed on firm footing and where further work is required.
Given that the discovery of wave DM could be imminent -- as a single example, ADMX is probing parts of the most well motivated axion parameter space right now~\cite{ADMX:2018gho,ADMX:2019uok,ADMX:2021nhd} -- these are fundamental problems to resolve.

Before beginning, let us outline how the discussion is organized.
We begin in Sec.~\ref{sec:density} by considering the possible form for the density matrix of DM.
Based on a quantum mechanical analog of the central limit theorem introduced by Glauber~\cite{Glauber:1963tx}, we argue that the density matrix most likely takes the form of a Gaussian weighting of coherent states.
In Sec.~\ref{sec:classical} we discuss when a quantum field description of DM can be replaced by a classical wave picture.
In particular, we emphasize that the field having large occupation, which occurs locally $m \ll 10\,\eV$, is only one of two conditions, the second being that the density matrix is sufficiently classical in a manner we define.
This second condition is satisfied by the Gaussian density matrix, but not in general.
When the classical wave picture applies, we show that ultralight DM formally behaves as a stochastic random variable, with statistical fluctuations inherited from the density matrix that can be passed to various experimental measurements.
For a Gaussian density matrix DM becomes a Gaussian random field.

In the two following sections we study the coherence properties of DM when treated as a classical wave.
We begin in Sec.~\ref{sec:frequency} by studying unequal spacetime correlators of the DM field -- in particular the autocorrelation function -- and its frequency domain analog, the power spectral density (PSD).
In Sec.~\ref{sec:coherence} we use the autocorrelation function and PSD to provide exact definitions of the DM coherence time, length, and volume.
Building on this, in Sec.~\ref{sec:g1g2} we extend the discussion to higher order measures of coherence that can be defined at the level of the quantum field rather than classical wave.

Removing the classical wave assumption, in Sec.~\ref{sec:boundary} we demonstrate that the formalism allows explicit computations to be performed through the wave-particle transition.
We focus on a specific calculation of the fluctuations of the DM energy density in a given volume, and show that at the scale of the coherence volume these fluctuations smoothly transition as we vary the DM mass from being Poisson distributed for particle-like DM to exponentially distributed for wavelike DM.
For DM with a mass near the wave-particle boundary, the fluctuations are neither Poisson nor exponential, but can be computed exactly.
Within volumes much larger than the coherence volume, the fluctuations become Gaussian, however the variance of the distribution depends explicitly on whether DM is in the wave or particle regime.

A key assumption underpinning many of our results is that the density matrix takes a Gaussian form.
We discuss deviations from this picture in Sec.~\ref{sec:NG}, and explain how these could be directly measured shortly after wave DM was discovered, as they would appear in the form of non-Gaussianities in the statistics of the field field.
Finally, in Sec.~\ref{sec:discussion} we discuss the various paths open to extend and formalize the quantum based approach to wave DM.

%%%%%%%%%%%%%%%%%%%%%%%%%%%%%%%
\section{The density matrix of dark matter}
\label{sec:density}
%%%%%%%%%%%%%%%%%%%%%%%%%%%%%%%

The fundamental description of a quantum system is provided by the density matrix and so we begin our discussion with a consideration of what form the density matrix should take for DM.
To do so, we draw inspiration from another field where a classical description was eventually replaced by a quantum analog: electromagnetism.
Indeed, many of the details in this section represent a translation of Glauber's foundational paper in quantum optics~\cite{Glauber:1963tx} to DM.

The experimental search for DM exploits standard model observables -- nuclear recoils in a detector, gamma-rays at a telescope, power in a cavity -- that could arise from a DM interaction.
Such observables are associated with Hermitian operators derived from the ultimate theory of DM-standard model interactions, within which the DM appears as a field operator.
In the case of scalar or pseudoscalar DM, the operator takes the form,\footnote{As the existence of quantum optics makes clear, the discussion can be generalized to DM with higher integer spin, for example dark photons.
Nevertheless, for simplicity we primarily focus on the (pseudo)scalar case throughout.}
\be
\hat{\phi}(t,\bx) = \int \frac{d^3\bk}{(2\pi)^3} \frac{1}{\sqrt{2\omega_{\bk}}} \left( \hat{a}_{\bk} e^{-i k \cdot x} + \hat{a}_{\bk}^\dag e^{i k \cdot x} \right)\!,
\label{eq:phihat}
\ee
with $k \cdot x = \omega_{\bk} t - \bk \cdot \bx$.
Here we are working in the interaction picture -- apparent from the presence of the time dependence in the operator -- and further we have assumed we are discussing a free field in flat space so that we can expand the operator in spatial plane waves.
(See Ref.~\cite{Kim:2021yyo} for an example of where the plane-wave expansion can be inappropriate even for direct detection.)

The fact that Eq.~\eqref{eq:phihat} is a quantum operator rather than a classical field is encoded in the creation and annihilation operators, which indicate that we are describing our field with an infinite set of harmonic oscillators, each of which satisfy the conventional commutation relation
\be
[\hat{a}_{\bk},\,\hat{a}_{\bq}^\dag] = (2\pi)^3 \delta(\bk-\bq).
\label{eq:com-cont}
\ee
Intuitively, the weighting of the various $\bk$ modes in $\hat{\phi}$ determines how energy is distributed in the field.
However, to fully describe the system we also need to know the state to ascribe each harmonic oscillator mode.

In general, then, we wish to know the density matrix for each mode $\bk$, which we denote by $\hat{\rho}_{\bk}$.
Let us focus on a specific mode of the system, leaving the $\bk$ dependence implicit, although we restore it shortly.
To determine the density matrix for this single mode we follow Glauber~\cite{Glauber:1963tx} and turn to the coherent state $\vt \alpha \ra$.
We recall that coherent states are eigenstates of the annihilation operator, $\hat{a} \vt \alpha \ra = \alpha \vt \alpha \ra$, which are described by a single number $\alpha \in \mathbb{C}$.
Coherent states have a number of useful properties.
Here we simply note that while not orthogonal, coherent states are complete,
\be
\la \beta \vt \alpha \ra = e^{-\tfrac{1}{2}(|\alpha|^2+|\beta|^2)+\alpha\beta^*},\hspace{0.4cm}
\frac{1}{\pi} \int d\alpha\, \vt \alpha \ra \la \alpha \vt = 1,
\ee
where $d\alpha = d(\operatorname{Re} \alpha)\, d(\operatorname{Im} \alpha)$.
In fact, what proves to be a key property of coherent states is that the set of $\vt \alpha \ra$ are not just complete, but overcomplete, such that any coherent state can be expanded in terms of all other states---in other words the coherent states are not linearly independent.
(Further, the decomposition of any state in terms of the coherent states is not unique.)
This allows the density matrix to be decomposed into a diagonal weighting of coherent states, given by
\be
\hat{\rho} = \int d\alpha\,P(\alpha) \vt \alpha \ra \la \alpha \vt.
\label{eq:GSP}
\ee
This remarkable result is known as the Glauber-Sudarshan representation~\cite{Glauber:1963tx,Sudarshan:1963ts}.
Although the coherent states are overcomplete, $P(\alpha)$ is unique and can be determined for a given $\hat{\rho}$ using the inversion formula derived by Mehta~\cite{mehta1967diagonal},
\be
P(\alpha) = \int \frac{d\beta}{\pi^2}\, \la - \beta \vt \hat{\rho} \vt \beta \ra e^{|\alpha|^2+|\beta|^2+2i\operatorname{Im}[\alpha \beta^*]},
\label{eq:Mehta}
\ee
with $\vt \beta \ra$ an additional coherent state.

At this stage, Eq.~\eqref{eq:GSP} simply trades an unknown density matrix for an unknown weighting $P(\alpha)$.
Yet $P(\alpha)$ satisfies a number of interesting properties.
The hermiticity and unit trace of $\hat{\rho}$ imply that $P(\alpha) \in \mathbb{R}$ and $\int d\alpha\,P(\alpha) = 1$, respectively.
This is suggestive that $P(\alpha)$ can be interpreted as a probability distribution, although this cannot be correct, as Eq.~\eqref{eq:GSP} would then imply any density matrix can be decomposed as a classical weighting of $\vt \alpha \ra$, which are highly classical quantum states.
In general $P(\alpha)$ can only be interpreted as a distribution; it can take negative values for certain $\alpha$ and be more singular than a $\delta$-function.
In detail, $P(\alpha)$ only satisfies the unit measure requirement of Kolmogorov's three axioms for a probability distribution.
Therefore $P(\alpha)$ is not a classical probability distribution, although it can be treated as a quasi-probability distribution in a sense we review below.

Most importantly for our purposes, $P(\alpha)$ has enough of the essential elements of a classical probability distribution that it obeys a quantum mechanical analog of the central limit theorem, as shown in Ref.~\cite{Glauber:1963tx} and reviewed in App.~\ref{app:Quantum-CLT}.
The assumptions for this quantum central limit theorem to hold are as follows.
The field of interest must arise from the superposition of $n$ similar yet independent fields, each having an associated Glauber-Sudarshan representation, $\pi_i(\alpha)$, with $i$ labelling the different fields.\footnote{To clarify the sense in which we mean for the systems to be combined, consider the following optical analog.
Imagine $n$ lasers that are each directed towards a common point where they overlap.
If the optical field of each laser is described by $\pi_i(\alpha)$, the question is to determine the $P(\alpha)$ that describes the total field in the region where they all overlap.}
We further require all systems to satisfy $\pi_i(\alpha)=\pi_i(|\alpha|)$, which as shown in Sec.~\ref{sec:classical} implies each field is stationary and homogeneous.
Under those conditions, in the limit $n \to \infty$, the density matrix for the combined field takes the form of a Gaussian weighting of coherent states,\footnote{We emphasize that it is the number of systems, given by $n$, that becomes large, not the quantity $N$ that appears in Eq.~\eqref{eq:GaussianRho}.
As shown in App.~\ref{app:Quantum-CLT}, we can have $n \to \infty$ with $N \ll 1$.}
\be
\hat{\rho} = \int d\alpha\,\frac{1}{\pi N} e^{-|\alpha|^2/N}\, \vt \alpha \ra \la \alpha \vt.
\label{eq:GaussianRho}
\ee
We return to the interpretation of $N$ shortly.
For now we simply note that the coherent states are weighted by a two-dimensional Gaussian with zero mean and a variance of $N/2$,\footnote{Situations where the Gaussian has non-zero mean are discussed in Sec.~\ref{sec:NG}.} and that the density matrix is explicitly mixed, with purity $\textrm{Tr}[\hat{\rho}^2]=(1+2N)^{-1} < 1$.

The key question is whether Eq.~\eqref{eq:GaussianRho} describes the density matrix for the modes of DM, as has been assumed elsewhere in the literature (see e.g. Ref.~\cite{Kim:2021yyo}).
The result holds in many situations within electromagnetism.
For example, if a system has thermalized, it has a density matrix in the Gaussian form, although this is not a necessary condition; Eq.~\eqref{eq:GaussianRho} is more general.
Heuristically, we could imagine that processes such as galaxy mergers and the violent relaxation associated with virialization~\cite{Lynden-Bell:1966zjv} are sufficient to render the local DM density a sum of independent distributions to which the central limit theorem applies.
Further, if the system ever thermalized during its evolution then the Gaussian form must hold.
Yet ultralight DM is not expected to have a thermal origin and given the weakness of its interactions, DM's density matrix may not be Gaussian.
We return to consider this question in detail in Sec.~\ref{sec:NG}, in particular we discuss various alternatives for $\hat{\rho}$ and demonstrate how the density matrix can imprint itself in experimental measurements, thereby opening a path to directly accessing it after the discovery of DM.
In particular, if DM were in a pure coherent state, $P(\alpha) = \delta(\alpha-\beta)$, its behavior differs considerably from that of the Gaussian distribution in a manner experiments could test.
Until then, we generally assume Eq.~\eqref{eq:GaussianRho} holds for each mode of the DM field and explore the consequences.
We note, however, the full Gaussian form is generally only required to determine the statistical fluctuations of observables, their mean value is often fixed with the weaker condition that $P(\alpha)=P(|\alpha|)$ (this is true of many of the coherence properties we study in Secs.~\ref{sec:frequency}, \ref{sec:coherence}, and \ref{sec:g1g2}).
We are explicit below regarding which results require a specific $\hat{\rho}$ and which are more general.

Equation~\eqref{eq:GaussianRho} alone is insufficient to specify the form of the density matrix, we also need to provide an interpretation for $N$.
At this stage it is convenient to restore the fact that we have a continuum of density matrices, one for each $\bk$, and therefore we need to specify $N_{\bk}$.
As the notation suggests, $N_{\bk}$ is the expected occupation number of the state in that mode, which can be confirmed explicitly,
\be
\la \hat{N}_{\bk} \ra = \textrm{Tr} [\hat{\rho}_{\bk}\, \hat{a}_{\bk}^\dag \hat{a}_{\bk}] = N_{\bk}.
\label{eq:Nexp}
\ee
The mean number of states in a given mode can be expressed in a frame-independent manner as the ratio of the density of particles to the density of states.
The density of states for a free field is given directly by ${\cal N}_s = g_s/(2\pi)^3$, where $g_s$ is the number of degrees of freedom associated with the field: one for a scalar, two for a massless vector or graviton, and three for a massive dark photon.\footnote{Although we include a general $g_s$ in the present discussion, we emphasize that our primary focus is scalar fields.
To extend beyond this, we would need to include a sum over polarizations in Eq.~\eqref{eq:phihat}.}
The density of field quanta is described by the phase space density, ${\cal N}(\bx,\bk,t)$.
In scenarios where the spatial distribution is stationary and homogeneous, as we often approximate to be the case locally for DM, we can write ${\cal N}(\bx,\bk,t) = \bar{n}\,p(\bk)$, where $\bar{n}$ is the number density and $p(\bk)$ is the momentum distribution of the field, for instance, in the case of DM we can approximate it using the standard halo model studied in Sec.~\ref{sec:frequency}.\footnote{In actuality the DM phase space is neither stationary nor homogeneous.
Effects such as daily and annual modulation or gravitational focusing induce time and position dependent shifts in ${\cal N}(\bx,\bk,t)$, as discussed in e.g. Refs.~\cite{Lee:2013wza,Foster:2017hbq,Foster:2020fln,Kim:2021yyo}.
However, as we discuss in the conclusion of Sec.~\ref{sec:coherence}, the question is the stability of the phase space (or even the density matrix as a whole) on scales over which the DM field is correlated: the coherence length and time.
For a wide range of DM masses this condition is satisfied.}
We then arrive at,
\be
N_{\bk} = \frac{(2\pi)^3}{g_s}\, \bar{n}\,p(\bk).
\label{eq:Nk}
\ee
Although this provides us a form of the variance for the Gaussian weighting in Eq.~\eqref{eq:GaussianRho}, the result is more general and gives the expected occupation number according to Eq.~\eqref{eq:Nexp} independent of the form of $\hat{\rho}_{\bk}$.

Physically $N_\bk$ represents the mean number of states in a given $\bk$ mode.
It therefore contains information regarding distinguishability: for $N_{\bk} > 1$ there is no quantum number or label that distinguishes these states and they therefore must be regarded as identical.
This purely bosonic phenomena where we can have a large number of identical states is intimately related to the wave-particle transition, with $N_{\bk} \gg 1$ playing a role in diagnosing when the classical field description becomes appropriate, as we show in Sec.~\ref{sec:classical}.
It is also related to the coherence volume, $V_c$, that is a measure of the physical volume occupied by the indistinguishable states.
We demonstrate this explicitly in Sec.~\ref{sec:coherence}.

In order to incorporate the information in Eq.~\eqref{eq:Nk} we need to generalize our description of the density matrix to multiple modes.
Whilst conceptually straightforward, let us briefly outline our notation for doing so, as we make regular use of it in what follows.
Especially for manipulations involving coherent states it is convenient to to discretize the modes by placing the scalar field in a box of volume $V$, in which case the field operator and commutation relations become,
\bea
\hat{\phi}(t,\bx) &= \sum_{\bk} \frac{1}{\sqrt{2 V \omega_{\bk}}} \left( \hat{a}_{\bk} e^{-i k \cdot x} + \hat{a}_{\bk}^\dag e^{i k \cdot x} \right)\!, \\
[\hat{a}_{\bk},\,\hat{a}_{\bq}^\dag] &= \delta_{\bk,\bq}.
\label{eq:phi-finiteV}
\eea
We use the discrete and continuum normalizations interchangeably below as convenient.
We review the details of how to move between the two normalizations in App.~\ref{app:contdisc}, as well as discussing the challenge of working exclusively with the continuum.
A general density matrix in the Glauber-Sudarshan $P$-representation is written as
\be
\hat{\rho} = \int d\{\alpha\}\,P(\{\alpha\}) \vt \{\alpha\} \ra \la \{\alpha\} \vt.
\label{eq:rho-multimode}
\ee
Here we have adopted the notation of Ref.~\cite{Mandel:1995seg}, where $\{\alpha\}$ denotes the set of all $\alpha_{\bk}$ modes.
For example, $\vt \{\alpha\} \ra = \prod_{\bk} \vt \alpha_{\bk} \ra$ and where the Gaussian weighting is appropriate, we take
\be
P(\{\alpha\}) = \prod_{\bk} \frac{1}{\pi N_{\bk}} e^{-|\alpha_{\bk}|^2/N_{\bk}}.
\label{eq:Gauss-multimode}
\ee

Before continuing to determine the implications of Eq.~\eqref{eq:Gauss-multimode} for the DM field, we briefly collect several additional properties of $P(\alpha)$.
The Glauber-Sudarshan representation has several advantages beyond the quantum central limit theorem.
If we have an operator that depends on creation and annihilation operators, we can evaluate its normally ordered expectation value as\footnote{By normal ordering, denoted by a pair of colons, we imply that in all expressions the creation operators are placed to the left of annihilation operators without accounting for the failure of individual terms to commute.}
\be
\la :\! \hat{A}(\hat{a},\hat{a}^\dag)\! : \ra
= \int d\alpha\, P(\alpha) A(\alpha),
\label{eq:opticalequivalence}
\ee
where on the right we have removed the hat from $A$ as it is now a classical function of the complex variable $\alpha$, equivalent to the operator but with $\hat{a}^{(\dag)} \to \alpha^{(*)}$.
This expression, which is often denoted as the optical equivalence theorem (see e.g. Ref.~\cite{Mandel:1995seg}) shows that we can rewrite quantum expectation values of normally ordered operators as seemingly classical expectation values, albeit over a quasi-probability distribution $P(\alpha)$.
A further unique property of $P(\alpha)$ is that its negativity is the signature of a quantum state, by which we mean a state whose predictions cannot be reproduced by any classical field (examples include a Fock state or squeezed state).
The intuition behind this result is that if $P(\alpha) \geq 0$ then the density matrix is a classical distribution over coherent states, which again are highly classical quantum states.
This particular advantage is unlikely to be relevant for DM: for quantum fields that couple weakly to experiment, measuring their quantum nature is a substantially more challenging prospect than detection (see e.g. Ref.~\cite{Carney:2023nzz}).
By contrast, $P(\alpha)$ has its drawbacks; it can be highly singular -- making certain calculations cumbersome -- and it is not well suited for evaluating expressions that are not normally ordered.
For this reason we discuss another convenient quasi-probability distribution, the Wigner distribution, in App.~\ref{app:wigner}.

%%%%%%%%%%%%%%%%%%%%%%%%%%%%%%%
\section{Wave Dark matter as a classical random field}
\label{sec:classical}
%%%%%%%%%%%%%%%%%%%%%%%%%%%%%%%

Armed with a density matrix, we return to the original focus of understanding the behavior of $\hat{\phi}(t,\bx)$ as it enters various experimental observables.
Technically, in doing so we must account for the quantum mechanical nature of the detector, which in general mandates the use of quantum measurement theory (see e.g. Ref.~\cite{Beckey:2023shi}).
We do not account for this here and instead focus solely on the quantum mechanics of DM.

Searches for wave DM generically consider couplings linear in the DM field, suggesting we should focus on the first moment of the field, $\la \hat{\phi}(t,\bx) \ra$.
Nevertheless, higher moments are important.
One often considers observables that are sensitive to the field squared -- such as the power deposited by the field -- or could even search for quadratic DM couplings, as done in Refs.~\cite{Beadle:2023flm,Banerjee:2022sqg}.
Further, higher moments inform the statistical properties of the observable; to provide an explicit example, for a Gaussian observable $A$, $\la A \ra$ determines the mean, whereas $\la A^2 \ra$ is needed to infer the variance.

With this motivation, we consider equal time and position correlators of the DM field, $\la \hat{\phi}^n(t,\bx)\ra$.
We start with $n=1$, and draw on the finite volume mode expansion in Eq.~\eqref{eq:phi-finiteV} and the general $P$-representation in Eq.~\eqref{eq:rho-multimode}.
Explicitly,
\bea
\la \hat{\phi}(t,\bx) \ra
= \sum_{\bk} \int \!d\alpha_{\bk}\,P(\alpha_{\bk})\sqrt{\frac{2}{V \omega_{\bk}}} \operatorname{Re} \left[ \alpha_{\bk} e^{-i k \cdot x} \right]\!.
\eea
For a general $P(\alpha_{\bk})$ this is non-zero, as the case of a pure coherent state makes clear.
It does, however, vanish for a broad class of weightings.
In particular, observe that the Gaussian expression in Eq.~\eqref{eq:Gauss-multimode} depends only on the magnitude of $\alpha_{\bk}$, $P(\alpha_{\bk}) = P(|\alpha_{\bk}|)$.
For any system dependent solely on the magnitude of $\alpha_{\bk}$, the integral over the phase of the coherent states dictates that $\la \hat{\phi}(t,\bx) \ra=0$; in fact, all odd moments of the field vanish.
The even moments need not and for the explicit case of the Gaussian $P(\alpha_{\bk})$ we can compute\footnote{This calculation is most easily performed using the Wigner distribution reviewed in App.~\ref{app:wigner}, see in particular Eq.~\eqref{eq:sym_operator}.}
\be
\la \hat{\phi}^{2n}(t,\bx) \ra = (2n-1)!! \left( \int \frac{d^3\bk}{(2\pi)^3} \frac{N_{\bk}+\tfrac{1}{2}}{\omega_{\bk}} \right)^{n}\!.
\label{eq:hatphi2n}
\ee
Here $(2n-1)!! = (2n)!/2^nn!$ so that we recognize these as the moments of a Gaussian distribution of mean zero and variance that can be read off from the expression when $n=1$.
We conclude that measurements of $\hat{\phi}(t,\bx)$ fluctuate according to a Gaussian distribution.

Equation~\eqref{eq:hatphi2n} needs to be interpreted carefully; the integral that appears in the expression is UV divergent and must be regulated.
The problem is a zero-point quadratic divergence, $\int^{\Lambda} d^3\bk/\omega_{\bk} \sim \Lambda^2$, which can be regulated in a field theoretic manner as we briefly discuss in App.~\ref{app:zeropoint}.
An identical divergence arises in quantum optics in correlators of the electric or magnetic field (see e.g. Ref.~\cite{Mandel:1995seg}) and, as in the field theoretic picture, the problem is resolved with a refined interpretation of what is being measured.
For instance, an experiment is only sensitive to the field over a smeared spacetime position or equivalently a finite set of frequencies, which is sufficient to regulate the UV divergence.

These initial calculations illustrate that with the density matrix we can compute arbitrary observables of the DM field directly without recourse to a classical wave picture.
We continue in this spirit in Sec.~\ref{sec:boundary} and demonstrate that one can explicitly compute observables for an arbitrary DM mass and track how they behave as we move between the wave and particle regimes.
For the moment, however, a heuristic interpretation of Eq.~\eqref{eq:hatphi2n} also suggests that a simplification occurs for $N_{\bk} \gg 1$.
We explore this below, and show that this is a necessary although not sufficient requirement for treating DM as a classical field.

To study the $N \gg 1$ limit, observe that when computing expectation values of the DM field using the coherent state expansion in Eq.~\eqref{eq:GSP}, in all normally ordered expressions we simply send $\hat{a}^{(\dag)} \to \alpha^{(*)}$.
Not all observables are normally ordered.
For example, the creation and annihilation operators in the moments of the field operator are symmetrically ordered, and involve expressions of the form $\hat{a}^{\dag} \hat{a} + \hat{a} \hat{a}^{\dag} = 2 \hat{N} + 1$ as is apparent from  Eq.~\eqref{eq:hatphi2n}.
When $N \gg 1$, however, we can approximate $[\hat{a}_{\bk},\,\hat{a}_{\bq}^\dag] \simeq 0$, trivially normal order operators of interest, and compute their expectation value as follows (cf. Eq.~\eqref{eq:opticalequivalence}),
\be
\la \hat{A}(\hat{a},\hat{a}^\dag) \ra
= \int d\alpha\, P(\alpha) A(\alpha) + {\cal O}(1/N).
\label{eq:largeN}
\ee

Applying this logic to the DM field, in limit where $N_{\bk}$ is large for all relevant modes, expectation values can be computed by replacing the creation and annihilation operators within $\hat{\phi}$ with the appropriate coherent states.
Accordingly, the field operator is approximately a c-number,
\be
\hat{\phi}(t,\bx) \simeq \phi(t,\bx) = 
\sum_{\bk} \sqrt{\frac{2}{V \omega_{\bk}}} \operatorname{Re} \left[ \alpha_{\bk} e^{-i k \cdot x} \right]\!.
\label{eq:phicnumber}
\ee
This is a clear simplification.
Nonetheless, we cannot treat $\phi$ as a classical background field: the $\alpha$ values that then appear within $\phi$ are resolved through integration against $P(\alpha)$ as in Eq.~\eqref{eq:largeN}, and in general $P(\alpha)$ can be highly singular and is not a probability distribution.
If $P(\alpha) \geq 0$, however, then Eq.~\eqref{eq:largeN} represents the calculation of a classical expectation value with $P(\alpha)$ the probability distribution function, which implies we can treat each $\alpha_{\bk}$, and therefore $\phi$, as a stochastic c-number.
Under these conditions, we are justified in treating the DM field operator $\hat{\phi}$ as a classical stochastic background field in all calculations; in other words, as a classical random wave.\footnote{Given that demonstrating $P(\alpha) < 0$ for DM is likely to be extremely challenging experimentally~\cite{Carney:2023nzz}, it may be that in practice the second requirement can be relaxed.}

We can gain intuition as to why $N \gg 1$ alone is insufficient for the classical wave picture to hold by considering a counterexample: the Fock state.
A state of definite particle number is a manifestation of the quantization of the field and cannot be reproduced by any combination of classical waves.
For a pure Fock state with $N$ particles the density matrix is $\rho = \vt N \ra \la N \vt$, which using Eq.~\eqref{eq:Mehta} implies
\be
P(\alpha) = \frac{e^{|\alpha|^2}}{N!} \partial_{\alpha}^N \partial_{\alpha^*}^N \delta(\alpha).
\label{eq:numberP}
\ee
Being more singular than a $\delta$-function $P(\alpha)$ must be negative within its domain and so exhibits the hallmark of a quantum state; a Fock state cannot be reproduced by a classical distribution of coherent states.
More to the point, there are measurements we can perform on the Fock state that a classical wave model would not reproduce.
As an explicit albeit contrived example, if one were able to directly measure the number of quanta in the DM field, such a measurement would record $N$ with zero variance.
While a classical wave could reproduce a mean number of $N$ counts, at best it can reduce the variance to the Poisson limit of $\sqrt{N}$ (see e.g. Refs.~\cite{Mandel:1995seg,Carney:2023nzz}).
This would then be an explicit measurement the classical wave picture would fail to reproduce.
Even though the Fock state is not classical in this sense, it can have arbitrarily large occupation and so is an example of where $N \gg 1$ does not imply the applicability of a classical wave picture.
This argument extends to other quantum states such as squeezed states.

To summarize, for $N_{\bk} \gg 1$, we can treat $\hat{\phi}$ as a c-number $\phi$ determined by $P(\alpha)$.
When $P(\alpha)$ is positive definite, $\phi$ behaves as a stochastic classical wave.
We can make clearer that this implies the usual wave picture of ultralight DM by considering a schematic form for the classical field.
Under these two conditions, we have from Eq.~\eqref{eq:Nexp} that $\la |\alpha_{\bk}| \ra \simeq \sqrt{N_{\bk}}$.
We then write $\alpha_{\bk} = |\alpha_{\bk}| e^{-i \varphi_{\bk}}$, so that Eq.~\eqref{eq:phicnumber} can be written schematically as
\be
\phi(t,\bx) \simeq
\sum_{\bk} \sqrt{\frac{2 N_{\bk}}{V \omega_{\bk}}} \cos (\omega_{\bk} t - \bk \cdot \bx + \varphi_{\bk}).
\ee
The result is schematic because $\la |\alpha_{\bk}| \ra \simeq \sqrt{N_{\bk}}$ represents only the average behavior; in general we must account for the full statistics of $\alpha_{\bk}$ as determined from $P(\alpha_{\bk})$.
Nevertheless, we note this form is similar to existing efforts to determine the statistical properties of wave DM, in particular see Refs.~\cite{Foster:2017hbq,Centers:2019dyn,Foster:2020fln,Dror:2021nyr,Lisanti:2021vij,Gramolin:2021mqv,Amaral:2024tjg,Flambaum:2023bnw}.
The only difference is the exact pattern of statistical fluctuations, which we return to shortly.
Pushing further, if we imagine the system is non-relativistic and sufficiently dominated by the single mode of its mass $\omega_{\bk} = m$, then we can take $\omega_{\bk} \simeq m$, $N/V = \rho/m$, so that at $\bx=0$ we have
\be
\phi(t) \simeq \frac{\sqrt{2 \rho}}{m} \cos (m t + \varphi).
\ee
Up to a phase this matches the commonly used wavelike description in Eq.~\eqref{eq:naive}.
Through this derivation we can see the various approximations that were required to arrive at the result, and the magnitude of the corrections to this picture, thereby achieving one of the core goals of the present work.

Having justified the general picture, we next study in more detail the specific behavior of the classical wave in the case of a Gaussian $P(\alpha)$.
As a first observation, we note that $P(\alpha) = P(|\alpha|)$ is sufficient to demonstrate that expectations values which we can determine from $\phi(t,\bx)$ are invariant under shifts in the origin in both time and space, so that the field is both stationary and homogeneous.
This follows as the fact the probabilities depend only on the magnitude of the coherent state implies that we are free to perform a mode-by-mode shift of $\alpha_{\bk} \to \alpha_{\bk} e^{i k \cdot x_0}$ and thereby arbitrarily adjusting the origin of the field in Eq.~\eqref{eq:phicnumber}.
Technically, $P(\alpha) = P(|\alpha|)$ is a sufficient but not necessary condition for the field to be stationary and homogeneous~\cite{Mandel:1995seg}, and as such we refer to this condition as \textit{strong stationarity}.
We make repeated use of this condition throughout, although we emphasize it is not true in general: a pure coherent state does not obey strong stationarity.
When it does hold, we can compute that the field is uncorrelated with its derivative,
\be
\la \phi \partial_t \phi \ra = \la \phi \nabla \phi \ra = 0.
\label{eq:phidphi-independence}
\ee

Turning to the Gaussian, the freedom to shift the phase allows us to write the field and its derivatives as follows,
\begin{align}
&\phi = \sum_{\bk} \sqrt{\frac{2}{V \omega_{\bk}}} \operatorname{Re} \left[ \alpha_{\bk} \right]\!,\hspace{0.4cm}
\partial_t \phi = \sum_{\bk} \sqrt{\frac{2\omega_{\bk}}{V}} \operatorname{Im} \left[ \alpha_{\bk} \right]\!, \nonumber \\
&\hspace{1.5cm}\nabla \phi = -\sum_{\bk} \bk \sqrt{\frac{2}{V \omega_{\bk}}} \operatorname{Im} \left[ \alpha_{\bk}\right]\!.
\label{eq:phidphi-modes}
\end{align}
From Eq.~\eqref{eq:Gauss-multimode}, each $\alpha_{\bk}$ is an independent Gaussian random field with uncorrelated real and imaginary parts.
As a sum over many Gaussians, we therefore conclude that the field and its derivatives are themselves independent Gaussian random fields.
The independence between the field and its derivative is simply a manifestation of the more general result in Eq.~\eqref{eq:phidphi-independence}, and we note that it was already observed that the field should be Gaussian in this case in Ref.~\cite{Kim:2023pkx}.
Again all three of these fields have zero mean, and so the statistics are determined by the variances.
For the field we have,
\bea
\textrm{Var}(\phi) = \bar{n}\int d^3\bk\, \frac{p(\bk)}{\omega_{\bk}} = \bar{n} \left\la \frac{1}{\omega} \right\ra\!,
\label{eq:varphi}
\eea
where we used Eq.~\eqref{eq:Nk} with $g_s=1$ and in the final step leave the $\bk$ dependence of $\omega$ implicit.
We emphasize that the expectation value appearing in the final expression, denoted $\la \cdot \ra$, represents a shorthand for an average over the probability distribution for $\bk$.
This should be carefully distinguished from the other usages of the same expression that we employ, for instance Eq.~\eqref{eq:phidphi-independence} refers to an average over the classical wave fluctuations, whereas Eq.~\eqref{eq:hatphi2n} represents a quantum mechanical expectation value.

Although Eq.~\eqref{eq:varphi} is more general, for axion DM we have $\bar{n} \simeq \rho/m$ and $\la 1/\omega \ra \simeq 1/m$, so that $\textrm{Var}(\phi) = \rho/m^2$.
Interpreting averages as over time, this is consistent with what we would predict from Eq.~\eqref{eq:naive}: $\la \phi \ra = 0$ and $\la \phi^2 \ra = \rho/m^2$.
For the derivatives,
\be
\textrm{Var}(\partial_t \phi) = \bar{n} \la \omega \ra,\hspace{0.3cm}
\textrm{Var}(\nabla \phi) = \bar{n} \la \bk^2/\omega \ra.
\ee
Again we can confirm the expected behavior for non-relativistic DM, $\textrm{Var}(\partial_t \phi) \simeq \rho$ and $\textrm{Var}(\nabla \phi) \simeq 0$.
We emphasize, however, that the above results are more general, and could be applied to other scalar distributions in the wavelike regime, such as the relativistic cosmic axion background studied in Ref.~\cite{Dror:2021nyr}.

As the field is a Gaussian, we can trivially recompute the full set of moments using the classical wave picture.
The field is has zero mean and a variance set by Eq.~\eqref{eq:varphi} so that odd moments vanish and even moments are given by
\bea
\la \phi^{2n}(t,\bx) \ra 
&= (2n-1)!! \left( \bar{n} \left\la \frac{1}{\omega} \right\ra \right)^{n} \\
&= (2n-1)!! \left( \int \frac{d^3\bk}{(2\pi)^3} \frac{N_{\bk}}{\omega_{\bk}} \right)^{n}\!.
\eea
In the second step we rewrote the result to allow a direct comparison with the full quantum result in Eq.~\eqref{eq:hatphi2n}.
Comparing the two we see explicitly that moment-by-moment the error incurred from using the classical approximation is ${\cal O}(1/N_{\bk})$, as claimed, although the exact ${\cal O}(1)$ coefficient depends on the specific renormalization scheme adopted, see App.~\ref{app:zeropoint}.

To further explore the implications of the field being normally distributed, it is convenient to introduce quadrature variables,
\be
\hat{X}_{\bk} = \frac{\hat{a}_{\bk}+\hat{a}^{\dag}_{\bk}}{\sqrt{2}},\hspace{0.5cm}
\hat{Y}_{\bk} = -i\frac{\hat{a}_{\bk}-\hat{a}^{\dag}_{\bk}}{\sqrt{2}},
\label{eq:XY-quad}
\ee
which satisfy $[\hat{X}_{\bk},\hat{Y}_{\bk}]=i$ and $\hat{X}_{\bk}^2+\hat{Y}_{\bk}^2 = 2 \hat{N}_{\bk}+1$.
When the classical wave picture applies we can take $\hat{X}_{\bk} \simeq \sqrt{2} \operatorname{Re}[\alpha_{\bk}]$ and $\hat{Y}_{\bk} \simeq \sqrt{2} \operatorname{Im}[\alpha_{\bk}]$.
The two quadratures are therefore independent Gaussian random fields, and from Eq.~\eqref{eq:phidphi-modes} directly related to the individual modes of $\phi$ and $\partial \phi$, respectively.
As each quadrature of the DM field is normally distributed, if one isolates a single quadrature the statistical properties must differ from a measurement sensitive to both.
A further consequence of the quadratures being Gaussian distributed is that the number of states in a given mode, $\hat{N}_{\bk} \simeq \tfrac{1}{2}(\hat{X}_{\bk}^2+\hat{Y}_{\bk}^2)$, is distributed as a $\chi^2$ distribution with two degrees of freedom; equivalently, it is exponentially distributed.
This should be compared to the expectation for particle DM, where $N_{\bk}$ is Poisson distributed, corresponding to simply counting the number of particles with a given $\bk$.
The exponential distribution has a larger variance than Poisson, which can be heuristically associated with constructive and destructive interference of the wave.
We establish that these two limits are smoothly connected in Sec.~\ref{sec:boundary}.

To isolate the statistical fluctuations of the fields from the physical scales involved, we define normalized quadratures $\hat{x}_{\bk} = \hat{X}_{\bk}/\sqrt{N_{\bk}} \simeq \sqrt{2/N_{\bk}} \operatorname{Re}[\alpha_{\bk}]$ and $\hat{y}_{\bk} = \hat{Y}_{\bk}/\sqrt{N_{\bk}} \simeq \sqrt{2/N_{\bk}} \operatorname{Im}[\alpha_{\bk}]$.
For a Gaussian density matrix in the classical limit $x_{\bk}$ and $y_{\bk}$ are standard normal distributions (zero mean and unit variance) that are independent of each other and for each $\bk$ mode.
Explicitly, $\la x_{\bk} x_{\bq} \ra = \la y_{\bk} y_{\bq} \ra = \delta_{\bk,\bq}$ and $\la x_{\bk} y_{\bk} \ra = 0$.
Using these variables we can write
\be
\phi \simeq \sum_{\bk} \sqrt{\frac{N_{\bk}}{V \omega_{\bk}}}\,x_{\bk} ,\hspace{0.4cm}
\partial_t \phi \simeq \sum_{\bk} \sqrt{\frac{\omega_{\bk} N_{\bk}}{V}}\,y_{\bk},
\label{eq:xk-yk}
\ee
with a similar expression holding for $\nabla \phi$.

The normalized quadrature notation is particularly convenient for determining the statistical properties of wavelike DM observables.
We make use of it several times in the next section, however, we can already provide an example.
Consider the energy density in the DM field, which can be determined as
\begin{align}
\rho &= \frac{1}{2} \left[ (\partial_t \phi)^2 + (\nabla \phi)^2 + m^2 \phi^2 \right] \label{eq:rho-xy} \\
&= \sum_{\bk,\bq} \frac{1}{2V} \sqrt{\frac{N_{\bk}N_{\bq}}{\omega_{\bk} \omega_{\bq}}} \left[ 
(\omega_{\bk} \omega_{\bq} + \bk \cdot \bq) y_{\bk} y_{\bq} + m^2 x_{\bk} x_{\bq} \right]\!. \nonumber
\end{align}
The average value is as expected,
\be
\la \rho \ra = \int \frac{d^3\bk}{(2\pi)^3}\, \omega_{\bk} N_{\bk} = \la \omega \ra \bar{n}.
\ee
Turning to the statistical fluctuations, Eq.~\eqref{eq:rho-xy} demonstrates that the density is the sum of two $\chi^2$ distributions, each with a single degree of freedom, but combined with different weightings.
In the non-relativistic limit, we have
\be
\rho \simeq \sum_{\bk,\bq} \frac{m}{2V}  \sqrt{N_{\bk} N_{\bq}} \left[ y_{\bk} y_{\bq} + x_{\bk} x_{\bq} \right]\!,
\label{eq:rho-stats}
\ee
so that $\la \rho \ra \simeq m \bar{n}$.
The weighting between the two quadratures is now identical, implying that the non-relativistic density is exponentially distributed, as studied in detail in Refs.~\cite{Kim:2023pvt,Kim:2023kyy}.
Away from the non-relativistic limit, the density is not exponentially distributed.

%%%%%%%%%%%%%%%%%%%%%%%%%%%%%%%
\section{Autocorrelation and the Frequency Domain}
\label{sec:frequency}
%%%%%%%%%%%%%%%%%%%%%%%%%%%%%%%

\begin{figure*}[!t]
\begin{center}
\includegraphics[scale=0.42]{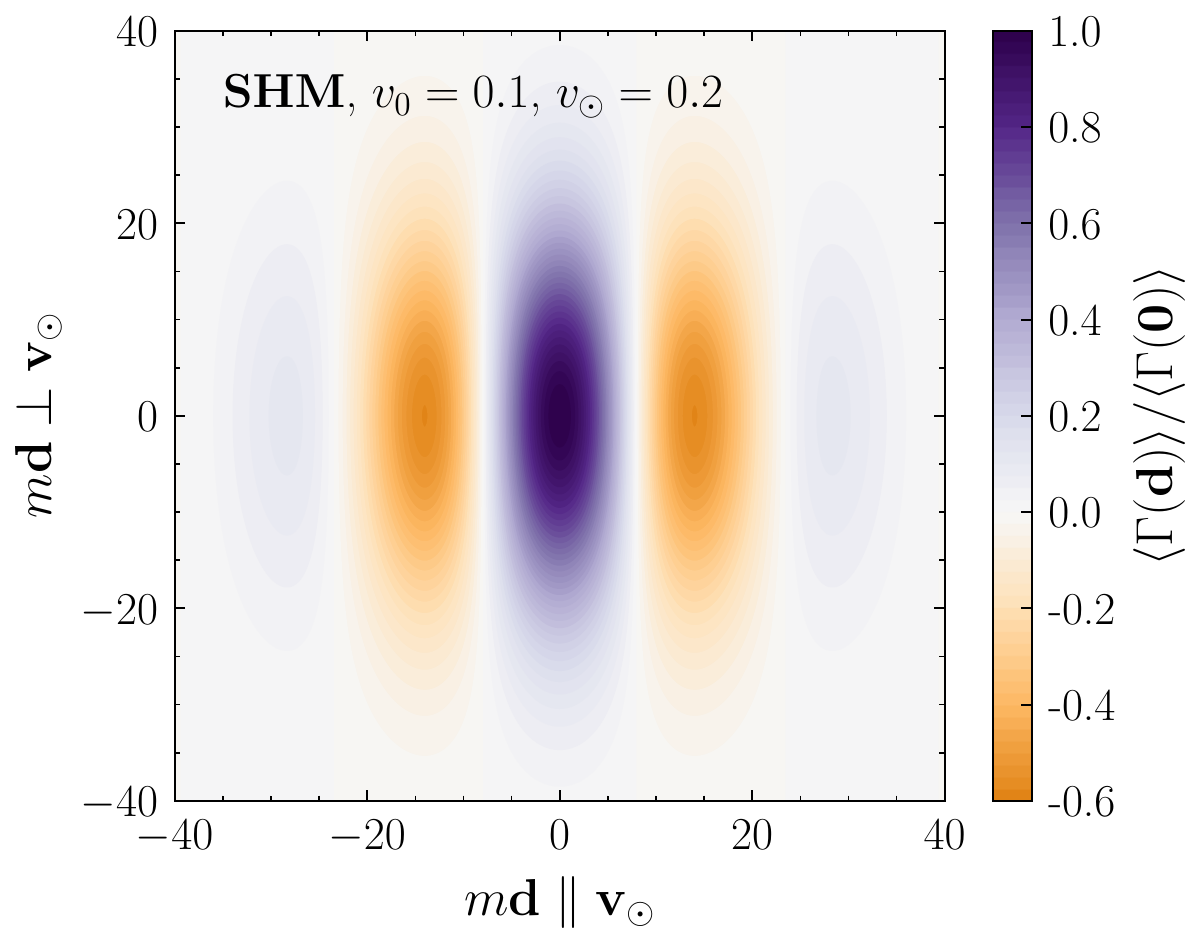}
\hspace{0.5cm}
\includegraphics[scale=0.42]{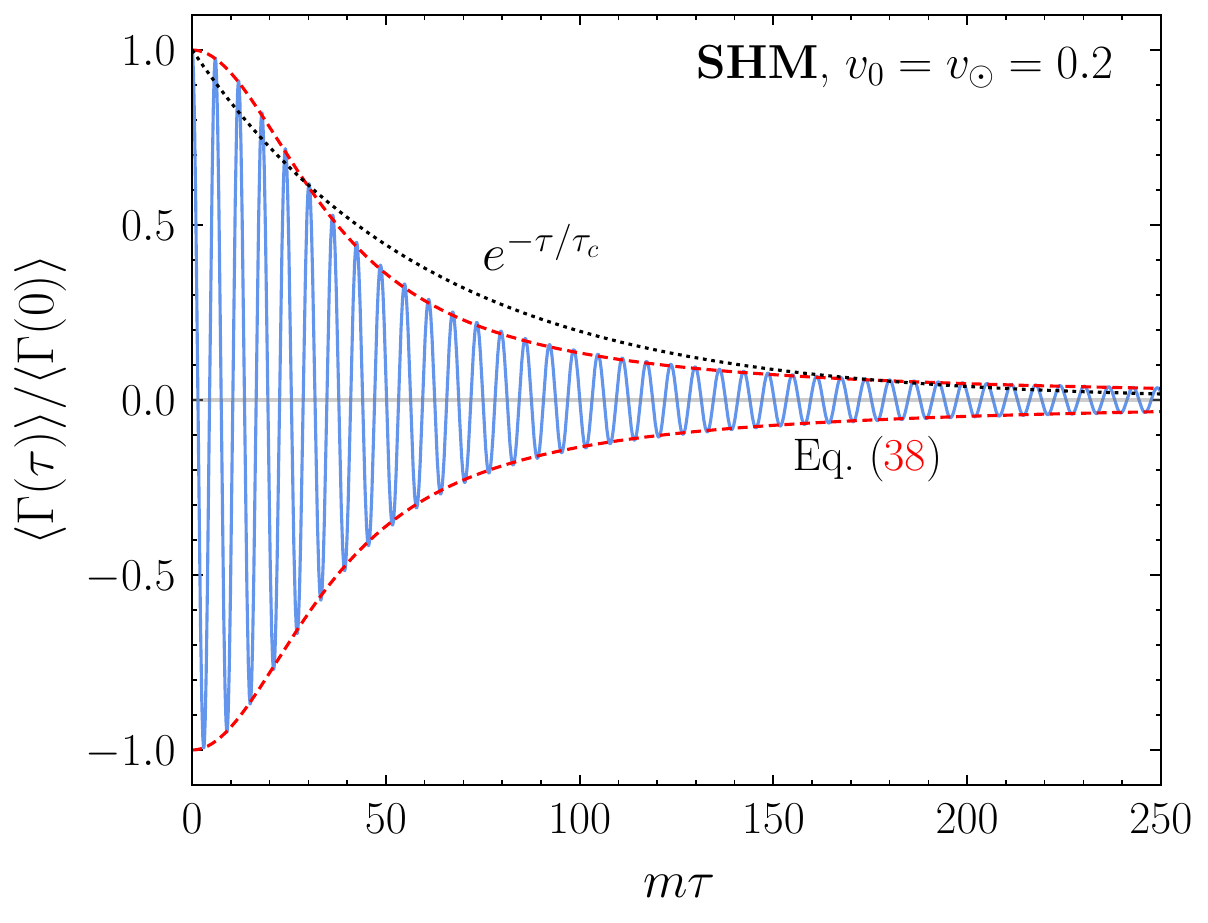}
\end{center}
\vspace{-0.7cm}
\caption{The DM spatial (left) and temporal (right) autocorrelation functions as defined in Eq.~\eqref{eq:auto}, assuming the standard halo model (SHM) as in Eq.~\eqref{eq:SHM} with the displayed parameters.
In addition to being oscillatory, both functions exhibit a decay that characterizes the coherence of the field as discussed in Sec.~\ref{sec:coherence}.
On the right we demonstrate that the envelope for the decay of the correlations is accurately represented by Eq.~\eqref{eq:Gt-SHM} (dashed red), although a simple exponential provides a rough approximation (dotted black).}
\label{fig:Gam-d-tau}
\end{figure*}

Thus far we have considered the properties of equal time and position correlation functions of the DM field.
In this section we turn to unequal spacetime correlators.
These objects provide a natural path to studying the field in the frequency domain as is commonly used in the analysis of wave DM experimental results.
They are also interesting objects in their own right.
The picture of wave DM invokes a sense in which the field can be correlated over potentially large time and distance scales, an idea that one often looks to exploit experimentally.
As we study in detail in Sec.~\ref{sec:coherence}, unequal spacetime correlators allow for this intuition to be rigorously quantified.

Consider the average autocorrelation function,
\be
\la \Gamma(\tau,\bd) \ra = \la \phi(t,\bx) \phi(t+\tau,\bx+\bd) \ra.
\label{eq:auto}
\ee
Intuitively, this is a measure of how knowledge of the field at a given spacetime position informs the field value at a separate position.
We emphasize three features of the way we have written $\Gamma(\tau,\bd)$.
Firstly, the fact it is independent of the position $(t,\bx)$ implies that we focus solely on stationary and homogeneous fields.
Secondly, as we have written the correlation function in terms of $\phi$ rather than $\hat{\phi}$, we are assuming the two classical wave conditions hold, $N_{\bk} \gg 1$ and $P(\alpha_{\bk}) \geq 0$.
A quantum analog is discussed in Sec.~\ref{sec:g1g2}.
Lastly, we focus on the average value of $\phi(t,\bx) \phi(t+\tau,\bx+\bd)$.
The statistical fluctuations can be computed similarly to how we studied the density fluctuations in Sec.~\ref{sec:classical}, and we consider them below.

In order to compute Eq.~\eqref{eq:auto}, we assume strong stationarity, $P(\alpha) = P(|\alpha|)$, which again the Gaussian distribution satisfies.
It then follows that,
\bea
\la \Gamma(\tau,\bd) \ra
= &\sum_{\bk,\bq} \frac{1}{2 V \sqrt{\omega_{\bk} \omega_{\bq}}} \\\times &\left\la \left( \alpha_{\bk} + \alpha_{\bk}^*\right)
\left( \alpha_{\bq} e^{-i q\cdot d} + \alpha_{\bq}^* e^{i q \cdot d} \right) \right\ra \\
= &\sum_{\bk} \frac{\la |\alpha_{\bk}|^2 \ra}{V \omega_{\bk}} \cos(k \cdot d),
\eea
where $d^{\mu} = (\tau,\bd)$.
From here, $\la |\alpha_{\bk}|^2 \ra = N_{\bk}$, which when combined with Eq.~\eqref{eq:Nk} leads to (cf. Ref.~\cite{Derevianko:2016vpm}),
\be
\la \Gamma(\tau,\bd) \ra = \bar{n} \int d^3\bk\, \frac{p(\bk)}{\omega_{\bk}} \cos(\omega_{\bk} \tau - \bk \cdot \bd).
\label{eq:Gamma-tau-d}
\ee
Observe that $\la \Gamma(\tau,\bd) \ra$ is an even function of all its arguments, which can also be derived as a consequence of stationarity and homogeneity.
Taking $\tau=0$ and $\bd=\mathbf{0}$ the result reduces exactly to Eq.~\eqref{eq:varphi}, as it must, however, if we only take $\bd=\mathbf{0}$ the result simplifies to\footnote{Although we use the same notation for both, the energy and 3-momentum distributions, $p(\omega)$ and $p(\bk)$, appearing in these expressions are not the same.
Appending subscripts to make the distinction manifest, the two are related by $p_{\omega}(\omega) = \int d^3\bk\,p_{\bk}(\bk) \delta(\omega-\omega_0)$, with $\omega_0^2 = k^2+m^2$.
For $p_{\bk}(\bk)=p_{\bk}(|\bk|)$ this simplifies to $p_{\omega}(\omega) = 4 \pi \omega k\, p_{\bk}(k)$.
A similar distinction must be made between the velocity and speed distributions in Eq.~\eqref{eq:Gam-DM}.}
\be
\la \Gamma(\tau) \ra = \bar{n} \int_0^{\infty} d\omega\, \frac{p(\omega)}{\omega} \cos(\omega \tau).
\label{eq:Gamma-tau}
\ee

In order to develop some intuition for these quantities, we consider several examples.
For a single mode, $p(\bk) = \delta(\bk-\bq)$, $\la \Gamma(\tau,\bd) \ra = (\bar{n}/\omega_{\bq}) \cos(\omega_{\bq} \tau - \bq \cdot \bd)$, which is an oscillatory function that does not decay for arbitrarily large $\tau$ or $\bd$.
For a single mode, knowledge of the field value at a single $(t,\bx)$ is sufficient to know it at all other positions and times.

Consider next non-relativistic DM, so that $\bk \simeq m \bv$ and $\omega_{\bk} \simeq m(1+v^2/2)$.
The temporal and spatial correlation functions are then approximately,
\bea
\la \Gamma(\bd) \ra &\simeq \frac{\bar{n}}{m} \int d^3\bv\, f(\bv) \cos(m \bv \cdot \bd), \\
\la \Gamma(\tau) \ra &\simeq \frac{\bar{n}}{m} \int dv\, f(v) \cos(m(1+v^2/2)\tau),
\label{eq:Gam-DM}
\eea
where $f(\bv)$ and $f(v)$ are the DM velocity and speed distributions, respectively, and we integrate over all values of $\bv$ and $v \geq 0$.
As a simple ansatz for these distributions, we can adopt the commonly assumed standard halo model (SHM)
\be
f(\bv) = \frac{1}{\pi^{3/2} v_0^3} \exp \left[\frac{-(\bv + \bv_{\odot})^2}{v_0^2} \right]\!,
\label{eq:SHM}
\ee
which is characterized by a dispersion $v_0$ and velocity of the Sun in the halo rest frame, $\bv_{\odot}$, conventionally taken as $v_0 \simeq 220\,\textrm{km/s}$ and $v_{\odot} \simeq 232\,\textrm{km/s}$, both ${\cal O}(10^{-3})$ in natural units.\footnote{Technically, $\bv_{\odot}$ in Eq.~\eqref{eq:SHM} should be replaced with the velocity of the observer, for instance at the location of an experiment on the surface of the Earth.
Corrections from the solar velocity are generally small, however, making the use of $\bv_{\odot}$ a good approximation in most cases as discussed in Ref.~\cite{Foster:2017hbq}.
In certain contexts, however, it is critical.
For two or more detectors the rotation of the Earth induces a large daily modulation effect that can be observed through DM interferometry as shown in Ref.~\cite{Foster:2020fln}.
How a time dependent $\bv_{\odot}$ can be included in our formalism is discussed in Sec.~\ref{sec:coherence}.}
The spatial autocorrelation function can be evaluated from this ansatz as,
\be
\la \Gamma(\bd) \ra \simeq \frac{\bar{n}}{m} \cos(m \bv_{\odot} \cdot \bd) e^{-(mv_0 d/2)^2},
\label{eq:Gd-SHM}
\ee
up to higher order corrections in $v_0,v_{\odot}$.
The function oscillates on a scale $d \sim 2\pi/mv_{\odot}$, but unlike for a single mode the correlations are exponentially damped at large distances.
We plot an example of this in Fig.~\ref{fig:Gam-d-tau} where the oscillations and decay can be observed.
The figure uses artificially enhanced values of the SHM parameters.
Without these the timescale for the correlations to decay becomes much larger than the scale of the natural oscillations, the inflated values are simply chosen to allow both scales to be observed.

The analytic form of the temporal correlation function is more complex, although it can be approximated as,
\be
\hspace{-0.19cm}\la \Gamma(\tau) \ra \simeq \frac{\bar{n}}{m} \frac{2\sqrt{2}\, \cos(m \tau)}{\big[4+(m v_0^2 \tau)^2 \big]^{3/4}} \exp \!\left[ - \frac{(m v_0 v_{\odot} \tau)^2}{4+(m v_0^2 \tau)^2} \right]\!.
\label{eq:Gt-SHM}
\ee
Beyond neglecting higher order terms in $v_0,v_{\odot}$, the additional approximation here is in the phase of the oscillations, which differs from $m \tau$ (we show how to derive the full result in Sec.~\ref{sec:g1g2}).
Regardless, the correlation function is again oscillatory, now on a scale $\tau \sim 2\pi/m$, and the correlations are damped at large times.
Both features can be seen for the normalized autocorrelation function in Fig.~\ref{fig:Gam-d-tau} for the exact (rather than approximate) result.
As studied in Sec.~\ref{sec:coherence}, the timescale for correlations to decay is related to the coherence time $\tau_c$ and a first approximation for the decay in the correlations is $e^{-\tau/\tau_c}$.
The figure also shows that Eq.~\eqref{eq:Gt-SHM} provides an accurate representation of the envelope of the decay.

The autocorrelation function further encodes the properties of the field in the frequency domain.
From the Wiener–Khinchin theorem, the PSD can be computed as\footnote{Although Eq.~\eqref{eq:PSD-def} is often more convenient, the average PSD can also be computed from the Fourier transform of $\phi(t)$, denoted $\tilde{\phi}(\omega)$, through $\la \tilde{\phi}^*(\omega) \tilde{\phi}(\omega') \ra = \pi \la S(\omega) \ra \delta(\omega-\omega')$.}
\be
\frac{1}{2} S(\omega) = \int_{-\infty}^{\infty} d\tau\,\Gamma(\tau) e^{i\omega \tau}.
\label{eq:PSD-def}
\ee
Using this and Eq.~\eqref{eq:Gamma-tau} we can immediately compute the average value, 
\be
\la S(\omega) \ra = \frac{2\pi \bar{n} p(\omega)}{\omega},
\label{eq:PSD}
\ee
which again assumes only strong stationarity.
For DM, the result is well approximated by (taking $\bar{n}=\rho/m$)
\be
\la S(\omega) \ra \simeq \frac{2\pi \rho f(v_{\omega})}{m^3 v_{\omega}},
\ee
where $v_{\omega} = \sqrt{2(\omega/m-1)}$.

Several comments regarding the PSD are warranted.
As $S(\omega)$ and $\Gamma(\tau)$ are Fourier transform pairs, the autocorrelation function being real and even in $\tau$ implies the PSD must be real and even in $\omega$.
Therefore, the PSD must have support for negative frequencies.
Yet as these frequencies carry no unique information (the PSD is an even function) and it is convenient to work with physical energies, the factor of $1/2$ in Eq.~\eqref{eq:PSD-def} is introduced so that we can work with a one-sided PSD.\footnote{Different conventions have been adopted in the axion literature, for instance this explains a relative factor of two between Eq.~\eqref{eq:PSD} and the equivalent result in Ref.~\cite{Dror:2021nyr}.
We discuss the various choices and their consistency in App.~\ref{app:PSDconv}.}
For example, this implies that we can use the PSD in Eq.~\eqref{eq:PSD} interpreting $p(\omega)$ to only have support for $\omega \geq 0$.
As an explicit example, we can consistently compute the total power in the field using only positive frequencies.
Integrating the average of Eq.~\eqref{eq:PSD-def}, we obtain
\be
\int \frac{d\omega}{2\pi}\,\la S(\omega) \ra = \la\Gamma(0)\ra = \la \phi^2(t) \ra = \bar{n} \la 1/\omega \ra,
\label{eq:plancherel}
\ee
which is consistent with Eq.~\eqref{eq:PSD}.
In the above expression and all that follow, integrals over frequency are interpreted as over $\omega \in [0,\infty)$, whereas integrals over time are evaluated for $\tau \in (-\infty,\infty)$, unless specified otherwise.

We end the section with a discussion of the statistical properties of $\Gamma(\tau,\bd)$ and $S(\omega)$.
For this purpose we again focus on the Gaussian $P(\alpha)$, although the discussion can be readily generalized to other distributions.
Using the normalized quadratures as in Eq.~\eqref{eq:xk-yk} the autocorrelation function can be written as,
\be
\Gamma(\tau,\bd) = \sum_{\bk,\bq} \frac{1}{V} \sqrt{\frac{N_{\bk} N_{\bq}}{\omega_{\bk}\omega_{\bq}}}\,x_{\bk} [x_{\bq} \cos(q \cdot d) + y_{\bq} \sin(q \cdot d)].
\label{eq:Gam-stats}
\vspace{0.1cm}
\ee
Consistency with Eq.~\eqref{eq:Gamma-tau-d} is apparent.
As $\Gamma(\tau,\bd)$ results from the product of correlated normal variables, the statistical properties can be determined using the known distribution from Ref.~\cite{NADARAJAH2016201}.
Taking the Fourier transform, the PSD is given by
\be
S(\omega) = \sum_{\bk,\bq} \frac{2\pi}{V} \sqrt{\frac{N_{\bk} N_{\bq}}{\omega_{\bk}\omega_{\bq}}}\,x_{\bk} [x_{\bq} + i y_{\bq}] \delta(\omega-\omega_{\bq}),
\ee
which again is consistent with Eq.~\eqref{eq:PSD}.
Explicit calculation and the repeated use of Wick's theorem demonstrates that,
\be
\la S^n(\omega) \ra = n! \la S(\omega) \ra^n.
\label{eq:Sn-exp}
\ee
Therefore for a Gaussian $P(\alpha)$ the PSD of the field is exponentially distributed, as argued heuristically for DM in Ref.~\cite{Foster:2017hbq} and for a more general field in Ref.~\cite{Dror:2021nyr} (cf. Eq.~\eqref{eq:rho-stats}).
It was further noted in Ref.~\cite{Foster:2017hbq} that the DC mode should differ to those with $\omega > 0$.
We can confirm this by returning to Eq.~\eqref{eq:Gam-stats},
\be
S(0) = \int d\tau\, \Gamma(\tau) = \sum_{\bk,\bq} \frac{2\pi}{V} \sqrt{\frac{N_{\bk} N_{\bq}}{\omega_{\bk} \omega_{\bq}}} \,x_{\bk} x_{\bq} \delta(\omega_{\bq}).
\ee
For the DC mode we have not included a factor of two as we did for the positive frequencies of the one-sided PSD, instead here the would-be positive and negative frequencies contribute equally at $\omega=0$.
We see the mode is indeed distributed differently, it behaves as a $\chi^2$-distributed with a single degree of freedom.
Accordingly, if $P(\alpha)$ takes a Gaussian form, then the likelihood frameworks developed for analyzing the data collected by wave DM experiments in Refs.~\cite{Foster:2017hbq,Foster:2020fln} are fully justified.

%%%%%%%%%%%%%%%%%%%%%%%%%%%%%%%
\section{Coherence of the classical field}
\label{sec:coherence}
%%%%%%%%%%%%%%%%%%%%%%%%%%%%%%%

A fundamental aspect of wave DM that enters into any experimental consideration of its detection is coherence.
In particular, the coherence time of the field, $\tau_c$, is a fundamental quantity in determining sensitivity of an experiment that measures the field for a time $T$.
There is a transition in the sensitivity scaling between $T < \tau_c$ and $T > \tau_c$ as the field loses coherence; for example, this limits how long DM can be resonantly excited until the power extracted saturates.
A common reference for this point in the DM literature is Ref.~\cite{Budker:2013hfa}; see also the discussion in Ref.~\cite{Dror:2022xpi}.
The transition is similar to that which occurs for counting experiment between a background free search and when there is an irreducible background, a prominent example of which is the neutrino fog for WIMP direct detection~\cite{Billard:2013qya,OHare:2021utq}.
The spatial coherence of the field, $d_c$, is another important ingredient in DM searches, as it dictates the distance within which independent detectors receive a correlated signal.
This can be exploited to perform interferometry on the DM wave or simply to enhance sensitivity by looking for a correlated signal over uncorrelated noise, see e.g. Refs.~\cite{Derevianko:2016vpm,Foster:2020fln,Chen:2021bdr,Masia-Roig:2022net}.

In spite of their central nature, these quantities are essentially always defined qualitatively as $\tau_c \sim 2\pi/m v^2$ and $d_c \sim (m v)^{-1}$, with $v \sim 10^{-3}$ (see e.g. Refs.~\cite{Graham:2013gfa,Foster:2020fln}).
Precise definitions can be provided, however, as these quantities are related to the scales over which the autocorrelation function decays, measuring the scales over which the field remains coherent as shown in Fig.~\ref{fig:Gam-d-tau}.
In this section we provide the relevant definitions and explore their properties.

We begin with the coherence time.
This has an established definition in the field of quantum optics (see e.g. Ref.~\cite{Mandel:1995seg}) in terms of the normalized autocorrelation function,\footnote{The coherence time is commonly defined using the complex analytic signal of $\phi(t)$ and we explain the connection to our definition in App.~\ref{app:PSDconv} (cf. also Ref.~\cite{Masia-Roig:2022net}).}
\be
\tau_c = 2 \int d\tau\,\left[ \frac{\la \Gamma(\tau) \ra}{\la \Gamma(0) \ra} \right]^2\!.
\label{eq:tauc-def}
\ee
As defined, $\tau_c$ encodes the temporal scale over which the autocorrelation function decays.
If dominated by a single mode, $\la \Gamma(\tau) \ra \sim \cos(\omega \tau)$, and the correlation time diverges.
If instead the autocorrelation function decays as $\la \Gamma(\tau) \ra \sim \cos(\omega \tau)\, e^{-|\tau|/\bar{\tau}_c}$ then assuming the field is sufficiently coherent (i.e. $\omega \bar{\tau}_c \gg 1$), we have $\tau_c \simeq \bar{\tau}_c$, cf. Fig.~\ref{fig:Gam-d-tau}.\footnote{The simple exponential decay model is not unique in identifying the coherence time.
Both $\la \Gamma(\tau) \ra \sim \cos(\omega \tau)\, \exp[-\pi \tau^2/2\bar{\tau}_c^2]$ and $\la \Gamma(\tau) \ra \sim \cos(\omega \tau) / [1+(\tau/\bar{\tau}_c)^2]^{3/4}$ (cf. Eqs.~\eqref{eq:Gd-SHM} and \eqref{eq:Gt-SHM}) also generate $\tau_c \simeq \bar{\tau}_c$, again assuming $\omega \bar{\tau}_c \gg 1$.}
We can define a dimensionless measure for the coherence of $\phi(t)$ by comparing $\tau_c$ to the mean oscillation period of the field $2\pi/\bar{\omega}$.
From these two quantities we construct a quality factor for the field (as in Ref.~\cite{Dror:2021nyr}),
\be
Q_{\phi} = \frac{\bar{\omega}\tau_c}{2\pi}.
\ee
The quality factor is a measure of how many periods the oscillations remain coherent, or more quantitatively, how many cycles it takes until $\la \Gamma(\tau)\ra \simeq \la\Gamma(0)\ra/e$.
It is usually assumed that for DM, $Q_{\DM} \sim 10^6$ so that the field remains coherent for many cycles and is therefore well approximated by Eq.~\eqref{eq:naive} for a long period, determined by $\tau_c$.
We confirm this intuition below.

Unless the average autocorrelation function is known, however, the definition in Eq.~\eqref{eq:tauc-def} is not particularly convenient.
It is therefore useful to find an alternative expression for the coherence time.
The inverse of Eq.~\eqref{eq:PSD-def} allows us to re-express the result in terms of the PSD,
\be
\tau_c = \frac{1}{\la \Gamma(0) \ra^2}
\int \frac{d\omega}{2\pi}\,\la S(\omega)\ra^2.
\ee
From here, using Eqs.~\eqref{eq:PSD} and \eqref{eq:plancherel} we arrive at,
\be
\tau_c = \frac{2\pi}{\la 1/\omega \ra^2} \int \frac{d\omega}{\omega^2}\, p(\omega)^2.
\label{eq:tc-pw}
\ee

Equation~\eqref{eq:tc-pw} lays bare that the coherence time is determined entirely by the energy distribution.
In particular, it is sensitive to the inverse width of $p(\omega)$ and diverges as the field becomes dominated by a single frequency.\footnote{Of course, in principle $p(\omega)$ could be arbitrarily complicated, such that in the most general case $\tau_c$ may only represent a crude measure of the evolution of the autocorrelation function.
Nonetheless, if the energy distribution is relative simple such that it is well characterized by its mean and standard deviation (as is the case for DM drawn from the SHM), then the coherence time represents an accurate measure as shown in Fig.~\ref{fig:Gam-d-tau}.}
To quantify these statements, consider the particularly simple scenario of a top-hat distribution -- $p(\omega) = (1/\delta\omega) \Theta(\omega-\bar{\omega}+\delta\omega/2) \Theta(\bar{\omega}+\delta\omega/2-\omega)$ -- with mean frequency $\bar{\omega}$ and width $\delta \omega$.
The coherence time is then,
\be
\tau_c = \frac{2\pi \delta \omega}{\bar{\omega}^2-(\delta \omega/2)^2} \ln^{-2}\left[\frac{\bar{\omega}+\delta \omega/2}{\bar{\omega}-\delta \omega/2} \right] \simeq \frac{2\pi}{\delta \omega},
\ee
where in the final step we assumed a narrow distribution, $\delta \omega \ll \bar{\omega}$.
As claimed, the coherence time diverges as $\delta \omega$ vanishes.

Turning to DM, the coherence time is determined by the speed distribution, and to leading order in $v \ll 1$ we obtain\footnote{The right of Eq.~\eqref{eq:tauDM} appeared repeatedly in Ref.~\cite{Foster:2017hbq} in the various analytic estimates that work provided for the sensitivity to axion DM, see for example their Eqs.~(45) and (55).
Our analysis clarifies that the appearance of that specific integral is because the coherence time plays a key role in the detectability of wave DM.
Indeed, the results in that work generically take the form of the sensitivity to axion couplings scaling as $\tau_{\DM}^{-1/4}$, the exact scaling argued for in Ref.~\cite{Dror:2022xpi}.}
\be
\tau_{\DM} \simeq \frac{2\pi}{m} \int \frac{dv}{v}\,f(v)^2.
\label{eq:tauDM}
\ee
For a simple top-hat model for the speed distribution with mean $\bar{v}$ and width $\delta v \ll \bar{v}$, we find the correlation time takes the expected form of $\tau_{\DM} \simeq 2\pi/m \bar{v} \delta v$.
A more interesting example is provided by the SHM in Eq.~\eqref{eq:SHM}.
Starting from Eq.~\eqref{eq:tc-pw}, we have
\bea
\tau_{\DM} =\, &\frac{\sqrt{2\pi} \operatorname{Erf} [\sqrt{2} v_{\odot}/v_0]}{m v_0 v_{\odot}} \\
\times &\left( 1 + \frac{3v_0^2}{4} - \frac{v_0 v_{\odot} e^{-2v_{\odot}^2/v_0^2}}{\sqrt{2\pi} \operatorname{Erf} [\sqrt{2} v_{\odot}/v_0]} + {\cal O}(v^4) \right)\!.
\label{eq:tDM-SHM}
\eea
Numerically, the second line is $\simeq 1 + 4 \times 10^{-7}$ and therefore even the ${\cal O}(v^2)$ corrections are completely irrelevant.
We have included the higher order terms to emphasize that in the approach we have outlined the coherence time is a rigorously defined quantity.
Numerically,
\be
\tau_{\DM} \simeq 2.80\,\textrm{ms}\, \left( \frac{1\,\mu\eV}{m} \right)\!.
\label{eq:tDM-Q}
\ee
Similarly, $Q_{\DM} = m \tau_{\DM}/2\pi \simeq 6.78 \times 10^5$.

Beyond quantitative numerical results, the explicit expressions also allow for the study of the coherence time's limiting behavior.
For instance, based on arguments similar to those in Ref.~\cite{Foster:2020fln}, one may expect the DM coherence time should take the form $\tau_{\DM} \sim 1/m v_0 v_{\odot}$, which is similar to Eq.~\eqref{eq:tDM-SHM}.
(We review those arguments at the end of this section.)
Nevertheless, the qualitative expression diverges as $v_{\odot} \to 0$, suggesting that an instrument at rest in the halo frame may have an enhanced sensitivity to wave DM.
This, of course, cannot be correct and our exact expressions reveal the appropriate behavior.
From Eq.~\eqref{eq:SHM}, when $v_{\odot} \to 0$, the speed distribution has a finite width and therefore should have a finite coherence.
Equation~\eqref{eq:tDM-SHM} manifests our expectation, as we have a finite coherence time of $\tau_{\DM} \simeq 4/m v_0^2$ as the observer's speed vanishes.
For $v_0 \to 0$, however, the width of the distribution vanishes and the coherence time should diverge, consistent with Eq.~\eqref{eq:tDM-SHM}.

As a final example, we can also confirm that the coherence time as defined above is a measure of how long the field remains well approximated by Eq.~\eqref{eq:naive}.
This intuition is embodied in the fluctuating phase model, where $\phi(t) = \phi_0 \cos(m t + \varphi(t))$, with $\varphi(t) \in [0,2\pi)$ a random phase that is resampled after a time that we now show can be identified as $\tau_c$.
This model is particularly convenient for time-domain analyses of wave DM, see e.g. Ref.~\cite{Dror:2022xpi}.
Given its natural definition in the time domain, it is more straightforward to compute the coherence time of this model from Eq.~\eqref{eq:tauc-def}.
To begin with, to ensure $\la \Gamma(\tau) \ra$ is symmetric we take the range of times where the phase is unchanged around $t=0$ as $t \in [-\bar{\tau}_c/2,\bar{\tau}_c/2]$ and for every interval of length $\bar{\tau}_c$ outside of this we sample a new random phase.
Consequently, $\la \Gamma(\tau) \ra/\la \Gamma(0) \ra = \cos(m \tau)\, \Theta(\tau+\bar{\tau}_c/2)\, \Theta(\bar{\tau}_c/2-\tau)$.
A direct computation then reveals that $\tau_c = \bar{\tau}_c + \sin(m \bar{\tau}_c)/m \simeq \bar{\tau}_c$, for $m \bar{\tau}_c \gg 1$.
Accordingly, the scale over which the fluctuating phase model jumps is exactly the coherence time of the field.

We next turn to consider the coherence properties of the field as a function of position.
Although the coherence time appears ubiquitously in considerations of the sensitivity to wave DM the coherence length and volume are equally fundamental concepts.
Indeed, as mentioned below Eq.~\eqref{eq:Nk}, the coherence volume is intimately related to $N_{\bk}$, which as we show in Sec.~\ref{sec:boundary} determines the transition between wave and particle behavior.
We therefore study first the coherence volume and establish the connection to $N_{\bk}$.
As for the coherence time, we can compute the volume in either of the conjugate variables, momentum or position, as
\be
V_c = \frac{(2\pi)^3}{\la 1/\omega \ra^2} \int \frac{d^3\bk}{\omega_{\bk}^2}\,p(\bk)^2 \simeq 
2 \int d^3\bd\,\left[ \frac{\la \Gamma(\bd) \ra}{\la \Gamma(\mathbf{0}) \ra} \right]^2\!.
\label{eq:Vc-def}
\ee
Unlike for the coherence time these two definitions are not equivalent; the first result should be taken as the definition, and the second an approximation.
We discuss the differences in App.~\ref{app:PSDconv}, however we note that for the SHM, the two agree at ${\cal O}(10\%)$.

As claimed in Sec.~\ref{sec:density}, the coherence volume can be thought of as the size of the region within which the bosonic particles cannot be distinguished.
We can see this as follows (a more detailed discussion can be found in Ref.~\cite{Mandel:1995seg}).
Consider the simple case where $p(\bk)$ is a narrow top hat of volume $(\delta k)^3$ centered at $\bk_0$, with $|\bk_0| \gg \delta k$.
Equation~\eqref{eq:Vc-def} then directly gives $V_c \simeq (2\pi)^3 p(\bk_0)$.
Multiplying this by the particle density $\bar{n}$, we obtain the number of particles in the coherence volume, $(2\pi)^3\, \bar{n}\,p(\bk_0)$, which when compared with Eq.~\eqref{eq:Nk} (taking $g_s=1$ for a scalar field with one degree of freedom), shows that $N_{\bk}$ can be thought of as the number of particles within the coherence volume.
Although this argument is heuristic, it reveals a fact consistent with the more realistic example considered below, which is that $V_c$ is an inverse measure of the width of $p(\bk)$.
As the momentum distribution narrows $V_c$ grows.
Hence, the number of indistinguishable states, $N_{\bk}$, grows also as momentum becomes a less informative label with which to distinguish them.

For DM the volume in Eq.~\eqref{eq:Vc-def} is well approximated by
\be
V_{\DM} \simeq \left( \frac{2\pi}{m} \right)^3 \int d^3\bv\,f(\bv)^2.
\ee
Turning again to the SHM for an explicit example, the volume takes the form,\footnote{If we instead computed the volume using Eq.~\eqref{eq:Gd-SHM} and the rightmost expression in Eq.~\eqref{eq:Vc-def}, the result is identical up to an overall factor of $1+e^{-2(v_{\odot}/v_0)^2} \simeq 1.1$, justifying the level of agreement quoted for the SHM.}
\be
V_{\DM} \simeq \left( \frac{\sqrt{2\pi}}{m v_0} \right)^3\!.
\label{eq:Vc-SHM}
\ee
As claimed, the coherence volume is sensitive to the width of the distribution, through $v_0$
Unlike the coherence time, it does not depend on the mean velocity, $-\bv_{\odot}$.
Further, the explicit form of the volume is suggestive that we can extract a coherence length of the form,
\be
d_{\DM} \simeq \frac{\sqrt{2\pi}}{m v_0} \simeq 674\,\textrm{m} \left( \frac{1\,\mu\eV}{m} \right)\!,
\label{eq:dDM-SHM}
\ee
and indeed this is what we would arrive at if we defined the coherence length in analogy to Eq.~\eqref{eq:tauc-def}, but as an integral over a single position rather than time.
For the SHM the width of $p(\bk)$ is independent of direction, and therefore we obtain an identical $d_{\DM}$ regardless of the direction the spatial correlations are measured.\footnote{If we measure correlations as in the rightmost expression in Eq.~\eqref{eq:Vc-def}, there is a small asymmetry in directions parallel and perpendicular to $\bv_{\odot}$.
However, as seen in Fig.~\ref{fig:Gam-d-tau}, these are simply due to the oscillations in the parallel direction rather than a change in the width, and when using the complex analytic signal to define coherence quantities, as we do in App.~\ref{app:PSDconv} and which gives rise to the left expression in Eq.~\eqref{eq:Vc-def}, such oscillations do not contribute.}
If, however, the width of $p(\bk)$ varies considerably between directions, the coherence length will vary equally as a function of direction.

Having defined the coherence length and time we note there is a heuristic connection between the two.
If we interpret $d_{\DM}$ as the spatial scale over which the DM wave is correlated, then the coherence time can be evaluated by considering how long one must wait for a new coherent patch to arrive at a given position.
For the SHM this is of the order $\tau_{\DM} \simeq d_{\DM}/v_{\odot}$, as argued in Ref.~\cite{Foster:2020fln}, and as accurately reflected by Eqs.~\eqref{eq:tDM-SHM} and \eqref{eq:dDM-SHM}.
Of course, these approximations can break down, such as when $v_{\odot} \to 0$.
In that limit, one can continue to interpret the result as above, but with the mean speed now replaced by $v_0$.
We emphasize, however, that the utility of the exact definitions in the present section is that we do not need to resort to heuristics: the coherence time, volume, and distance are instead exactly defined and interpreted as the scales over which the DM wave becomes uncorrelated.

Before moving on, let us return to our assumption of stationarity and homogeneity.
The results of this section and those in Sec.~\ref{sec:frequency} all required the strong stationarity condition of $P(\alpha) = P(|\alpha|)$, which again is a sufficient although not necessary requirement for expectation values to be independent of position and time.
For DM, however, this assumption must eventually break.
Daily and annual modulation induce a time dependence in $f(\bv)$ as the velocity of any Earthbound experiment changes in the halo frame.
Further, gravitational focusing can induce a change in the DM density throughout the year.
In both cases, the phase space and all the quantities we compute from it must vary which seemingly breaks our original assumption.
The key question, however, is how these quantities vary compared to the scales over which the DM becomes incoherent.
For instance, if the phase space varies much more slowly than the coherence time, then as the DM is effectively rendered incoherent after each $\tau_c$, we can recompute quantities within each coherence time interval using the phase space that holds within the appropriate time period.
Explicit calculations for how the field loses coherence over scales larger than the coherence time and volume are performed in Sec.~\ref{sec:boundary}.

This point is even more general than a spacetime dependency of the phase space.
Even if the Gaussian form of $P(\alpha)$ holds for DM locally, if DM did not have a thermal origin, the Gaussian form may not have held early on, implying at minimum there would be a time dependence to $\hat{\rho}$ over cosmological times as it evolves towards a Gaussian form (cf. App.~\ref{app:Pa-Evo}).
However, so long as the variation of the density matrix occurs over scales larger than $\tau_c$ and $d_c$, we can again generate reliable predictions for the fields behavior even for $t \gg \tau_c$ and $d \gg d_c$.
In particular, within each coherence time and volume we use the formalism introduced so far, performing calculations with the density matrix that holds for that spacetime region, and then smoothly glue the predictions together.

If we consider the explicit values of the relevant scales, then over the range of masses being probed for the QCD axion it is likely an excellent approximation that the field is stationary; using Eq.~\eqref{eq:tDM-Q}, even at the end of the QCD axion band (where $f_a \simeq M_\textrm{Pl}$) we have $m \simeq 10^{-12}\,\eV$ and hence $\tau_{\DM} \simeq 47\,\textrm{min}$.
This is sufficiently shorter than a day that accounting for daily modulation should prove no issue.
Similarly, at this mass $d_{\DM} \simeq 5\,\textrm{mAU}$, a scale over which the density should be constant even in the presence of gravitational focusing~\cite{Kim:2021yyo}.
If, however, we study wave DM with masses below the QCD window, stationarity and homogeneity are eventually violated: pushing towards the fuzzy DM regime~\cite{Hu:2000ke,Hui:2016ltb}, we have $\tau_{\DM} \simeq 10$ thousand years and $d_{\DM} \simeq 2\,$pc for $m \simeq 10^{-20}\,\eV$.
Nevertheless, for fuzzy DM masses the natural scales associated with DM are significantly larger than those of the Solar System that determine the variation of the local phase space.
Accordingly, from an effective field theory perspective, it seems likely that DM could not resolve Solar System level variations and therefore should depend only on appropriately averaged quantities.
Even if this is the case, there are still intermediate masses where the phase space varies on scales comparable to $\tau_{\DM}$ and $d_{\DM}$, where the results we have presented should be revisited.

%%%%%%%%%%%%%%%%%%%%%%%%%%%%%%%
\section{Higher Order Coherence}
\label{sec:g1g2}
%%%%%%%%%%%%%%%%%%%%%%%%%%%%%%%

In the previous two sections we extensively discuss the two-point correlation function of DM, its associated coherence properties, and formalized the connection with the coherence time and length used throughout the wave DM literature.
In quantum optics, rather than the autocorrelation function, it is common to study Glauber's $n$-point correlation functions $g^{(n)}$~\cite{Glauber:1963tx}, where $g^{(1)}$ is associated with two point correlations and for $n > 1$ higher order correlations are probed.
The most commonly studied functions are $g^{(1)}$ and $g^{(2)}$, which we define below, and indeed both have already been considered in the DM literature.
In this section we briefly review these higher order correlations and demonstrate that for a Gaussian $P(\alpha)$, these functions carry no additional information beyond the autocorrelation function studied in detail above.
Nevertheless, unlike $\Gamma(\tau,\bd)$, the $g^{(n)}$ correlators are defined in terms of quantum rather than classical fields so that we can relax the use of the classical wave approximation.

To define the correlation functions we first introduce a decomposition of the scalar field operator in Eq.~\eqref{eq:phi-finiteV} into positive and negative frequency modes,
\be
\hat{\phi}^+(x) = \sum_{\bk} \frac{1}{\sqrt{2 V \omega_{\bk}}}\hat{a}_{\bk} e^{-i k \cdot x},
\hspace{0.3cm}
\hat{\phi}^-(x) =  [\hat{\phi}^+(x)]^{\dag},
\label{eq:phi-plusminus}
\ee
such that $\hat{\phi}(x) = \hat{\phi}^+(x) + \hat{\phi}^-(x)$.
From these we define the first and second order coherence functions as,
\bea
g^{(1)}(x_1, x_2) &= 
\frac{\la \hat{\phi}^{-} (x_1)\hat{\phi}^{+} (x_2) \ra}
{\sqrt{\la \hat{\phi}^{-} (x_1) \hat{\phi}^{+} (x_1) \ra
\la \hat{\phi}^{-} (x_2)\hat{\phi}^{+} (x_2)\ra}}, \\
g^{(2)}(x_1, x_2  )  &=  \frac{\la \hat{\phi}^{-} (x_1) \hat{\phi}^{-} (x_2 )\hat{\phi}^{+} (x_2)\hat{\phi}^{+} (x_1) \ra}
{\la \hat{\phi}^{-} (x_1) \hat{\phi}^{+} (x_1) \ra 
\la \hat{\phi}^{-} (x_2) \hat{\phi}^{+} (x_2) \ra}.
\label{eq:def-g1g2}
\eea
The definition for higher order coherence functions follows similarly.
Observe that all expectation values are normal ordered and therefore primed to be evaluated using Eq.~\eqref{eq:opticalequivalence} and the $P(\alpha)$ quasi-probability distribution.
Both functions can be studied in general, but if we assume the field is stationary and homogeneous, then they only depend on $x^{\mu}_2-x^{\mu}_1=d^{\mu}=(\tau,\bd)$.

Consider $g^{(1)}(\tau,\bd)$ first.
If we assume strong stationarity as throughout Secs.~\ref{sec:frequency} and \ref{sec:coherence}, an identical calculation to the determination of $\la \Gamma(\tau,\bd) \ra$ in Eq.~\eqref{eq:Gamma-tau-d} yields,
\be
g^{(1)}(\tau,\bd) = \frac{1}{\la 1/\omega \ra} \int d^3\bk\, \frac{p(\bk)}{\omega_{\bk}} e^{-i k \cdot d},
\label{eq:g1pk}
\ee
which for $\bd=\mathbf{0}$ matches Ref.~\cite{Bernal:2024hcc}.
It also bears a striking resemblance to $\la \Gamma(\tau,\bd) \ra$, in particular,
\be
\operatorname{Re}\big[g^{(1)}(\tau,\bd)\big] = \frac{\la \Gamma(\tau,\bd) \ra}{\la \Gamma(0,\mathbf{0}) \ra}.
\label{eq:g1-2-Gam}
\ee
This is not an accident: $g^{(1)}(\tau,\bd)$ is the normalized autocorrelation function of the complex analytic signal, whereas $\la \Gamma(\tau,\bd) \ra/\la \Gamma(0,\mathbf{0}) \ra$ is the normalized autocorrelation function of the field itself.
Further details are provided in App.~\ref{app:PSDconv}.
Specifics aside, Eq.~\eqref{eq:g1-2-Gam} already suggests that we can obtain the coherence time and volume from the first order coherence function, and indeed we have
\be
\tau_c = \int d\tau\, \big|g^{(1)}(\tau,\mathbf{0})\big|^2,
\hspace{0.4cm}
V_c = \int d^3\bd\, \big|g^{(1)}(0,\bd)\big|^2,
\label{eq:tcVc-g1}
\ee
where as noted above we integrate over $\tau \in (-\infty,\infty)$.

Taking the non-relativistic limit appropriate for DM simplifies Eq.~\eqref{eq:g1pk} to,
\be
g^{(1)}(\tau,\bd) \simeq e^{-i m \tau}\int d^3\bv\, f(\bv) 
e^{i m(\bv \cdot \bd- v^2 \tau/2)}.
\label{eq:g1DM}
\ee
For the SHM of Eq.~\eqref{eq:SHM} we can compute this explicitly, finding
\begin{align}
\,&g_{\DM}^{(1)}(\tau, \bd) = \frac{e^{-im\tau}}{(1 + \xi^2)^{3/4}} \exp\left[- \frac{\left( v_0\bzeta + \bv_{\odot} \xi\right)^2}{v_0^2(1 + \xi^2)} \right]
\label{eq:g1SHM}\\
&\times \exp \left[- i \frac{\xi \left( (v_{\odot}/v_0)^2- \zeta^2 \right) + m \bv_{\odot} \cdot \bd}{1+\xi^2} - i \frac{3}{2} \arctan \xi \right]\!, \nonumber
\end{align}
where ${\boldsymbol \zeta} = \frac{1}{2}m v_0 \bd $, $\xi = \frac{1}{2} m v_0^2 \tau$.
This result matches that in Ref.~\cite{Blinov:2024jiz} up to a small difference in the exponential suppression (cf. Ref.~\cite{Derevianko:2016vpm}).
From this expression, $\la \Gamma(\bd) \ra$ and $\la \Gamma(\tau) \ra$ for the SHM follow immediately using Eq.~\eqref{eq:g1-2-Gam}, allowing us to confirm Eqs.~\eqref{eq:Gd-SHM} and~\eqref{eq:Gt-SHM}, including corrections to the phase of the oscillations for the latter.
We can further confirm the consistency of the result by computing
\begin{align}
\big|g_{\DM}^{(1)}(\tau, \mathbf{0})\big|^2 &= \frac{8}{\big[4+(m v_0^2 \tau)^2\big]^{3/2}}
\exp \left[ - \frac{2 (m v_0 v_{\odot} \tau)^2}{4+(m v_0^2 \tau)^2} \right]\!,\nonumber\\
\big|g_{\DM}^{(1)}(0, \bd)\big|^2 &= e^{- (m v_0 d)^2/2}.
\label{eq:g12-SHM}
\end{align}
Using Eq.~\eqref{eq:tcVc-g1} we can directly confirm these reproduce the expressions for the coherence volume in Eq.~\eqref{eq:Vc-SHM} and the leading order expression for the coherence time in Eq.~\eqref{eq:tDM-SHM}.
(The higher order contributions arise from the corrections to $\la 1/\omega \ra \omega_{\bk} \simeq 1$ and therefore to the higher order corrections to Eq.~\eqref{eq:g1DM}.)

The second order coherence function has a venerable history in quantum optics, given its association with the Hanbury Brown and Twiss effect~\cite{1954PMag...45..663B,1956Natur.177...27B,1956Natur.178.1046H}.
$g^{(2)}(\tau,\bd)$ can further diagnose the presence of genuine quantum behavior.
From Eq.~\eqref{eq:def-g1g2}, if we treat $\hat{\phi}$ as a c-number rather than a quantum field, then $g^{(2)}(0,\mathbf{0}) \geq 1$.
Nevertheless, for certain systems with $P(\alpha) < 0$ such as a Fock state, the correlation function can take a value in the classically forbidden range $0 \leq g^{(2)}(0,\mathbf{0}) < 1$.
(For a single mode Fock state with $N$ quanta, we have $g^{(2)}(\tau,\bd) = 1-1/N$.)
For a Gaussian density matrix, we instead have $P(\alpha) \geq 0$, and this does not occur.
Indeed, in this case the two correlation functions are related by~\cite{siegert1943fluctuations, 1999PhRvA..59.4595N}
\be
g^{(2)}(\tau,\bd) = 1+ \big|g^{(1)}(\tau,\bd)\big|^2.
\ee
Accordingly,
\bea
g^{(2)}(\tau,\bd) 
=&\, 1 + \left| \frac{1}{\la 1/\omega \ra} \int d^3\bk\, \frac{p(\bk)}{\omega_{\bk}} e^{-i k \cdot d} \right|^2 \\
\simeq&\, 1 + \left| \int d^3\bv\, f(\bv) 
e^{i m(\bv \cdot \bd- v^2 \tau/2)} \right|^2\!,
\eea
where the final line holds for a non-relativistic system such as DM.
In either event, we have $g^{(2)}(0, \mathbf{0}) = 2 > 1$.
Taking $\bd=\mathbf{0}$ the first line again matches Ref.~\cite{Bernal:2024hcc}.
To obtain an explicit example for the SHM, from Eq.~\eqref{eq:g1SHM} we obtain
\be
\hspace{-0.09cm}g_{\DM}^{(2)}(\tau, \bd) = 1 + \frac{1}{(1 + \xi^2)^{3/2}} \exp \! \left[- \frac{2\left( v_0\bzeta + \bv_{\odot} \xi\right)^2}{v_0^2(1 + \xi^2)} \right]\!.
\ee

%%%%%%%%%%%%%%%%%%%%%%%%%%%%%%%
\section{Description at the wave-particle boundary}
\label{sec:boundary}
%%%%%%%%%%%%%%%%%%%%%%%%%%%%%%%

We now explore the consequences of relaxing the assumption of large occupation to study the behavior of DM as the wave approximation breaks down.
In particular, we perform a calculation for arbitrary $N$ and demonstrate that there is a smooth transitions between the expected wave and particle behavior.
The calculation also reveals that there is an intermediate regime around $N \sim 1$ where neither the wave nor particle picture is fully appropriate.

Firstly, though, based on the discussions above, we can describe more accurately where we expect the wave-particle transition to occur.
In particular, combining the discussion of the coherence volume from Sec.~\ref{sec:coherence} with Eq.~\eqref{eq:Nk}, the transition is controlled by the number of indistinguishable particles per coherence volume, $N = \bar{n} V_c/g_s = \rho V_c/m g_s$, where the final expression holds for DM with a local energy density $\rho$.
Here $g_s$ enters as the internal degrees of freedom provide additional labels whereby the states can be distinguished.
Further, $V_c$ is determined by the velocity distribution and so we need to assume a form for this to compute $N$.
Taking the SHM and Eq.~\eqref{eq:Vc-SHM}, the number of indistinguishable particles is given by
\be
N = \frac{\rho \,(2\pi)^{3/2}}{g_s\,m^4v_0^3} \simeq \frac{1.22 \times 10^{29}}{g_s} \left( \frac{1\,\mu\eV}{m} \right)^4\!,
\label{eq:N}
\ee
or rearranging for $m$
\be
m = \left[ \frac{\rho \,(2\pi)^{3/2}}{g_sNv_0^3} \right]^{1/4} \simeq \frac{18.7\,\eV}{(g_sN)^{1/4}}.
\ee
Both numerical values above assume $\rho = 0.4\,\textrm{GeV/cm}^3$ and $v_0 = 220\,\textrm{km/s}$.
From the second result, we can read off the location of the wave-particle boundary by taking $N=1$; for an axion the transition occurs at $\sim$18.7\,eV, whereas for a dark photon (with $g_s=3$), the equivalent value is $\sim$14.2\,eV.
Both of these are consistent with the heuristic estimate of 10\,eV from the outset.\footnote{The result is comparable to other descriptions of the boundary in the literature.
The recent SNOWMASS reports defined the boundary at 1\,eV~\cite{Jaeckel:2022kwg,Adams:2022pbo}.
Other reviews have adopted a larger value, e.g. 30\,eV in Ref.~\cite{Hui:2021tkt}.}

\begin{figure*}[!t]
\begin{center}
\includegraphics[scale=0.28]{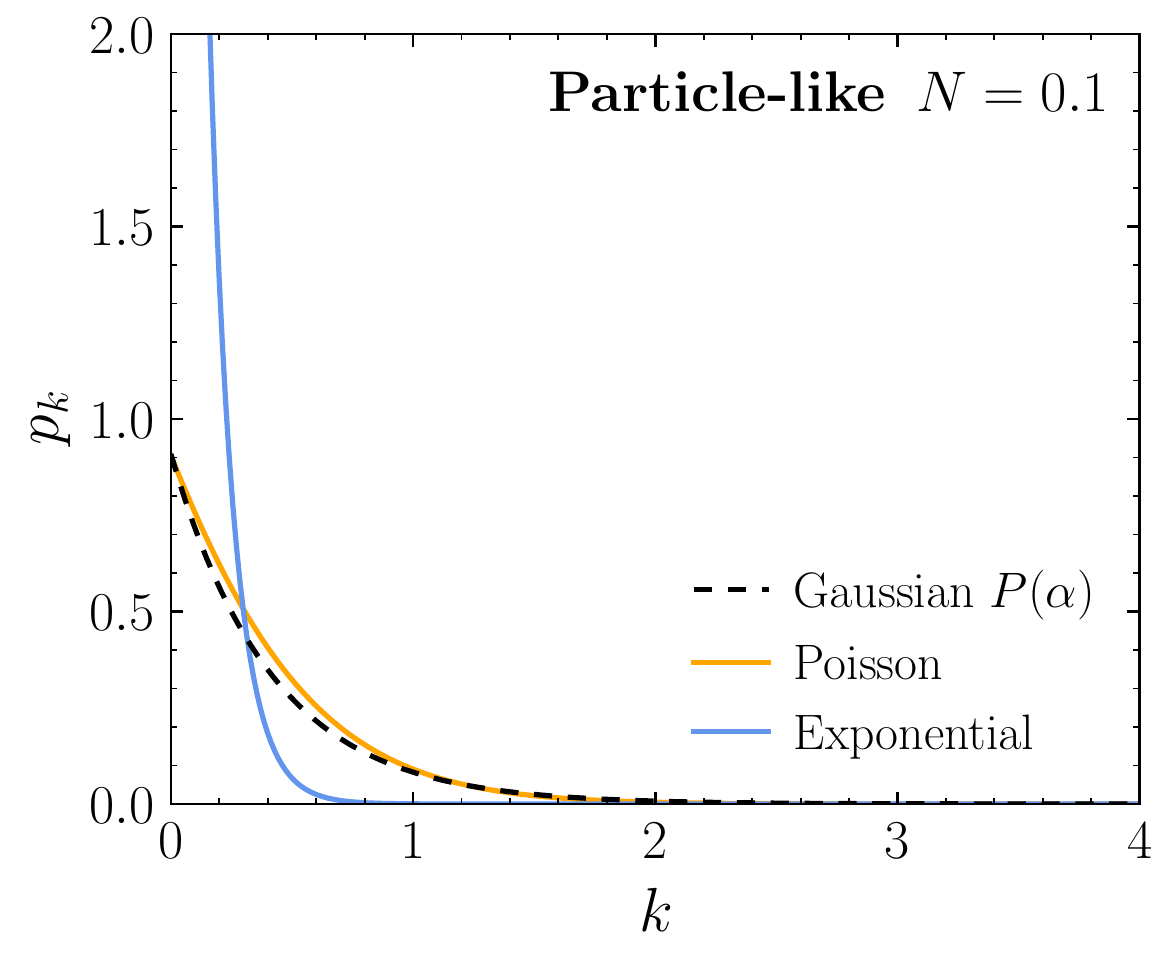}
\hspace{0.3cm}
\includegraphics[scale=0.28]{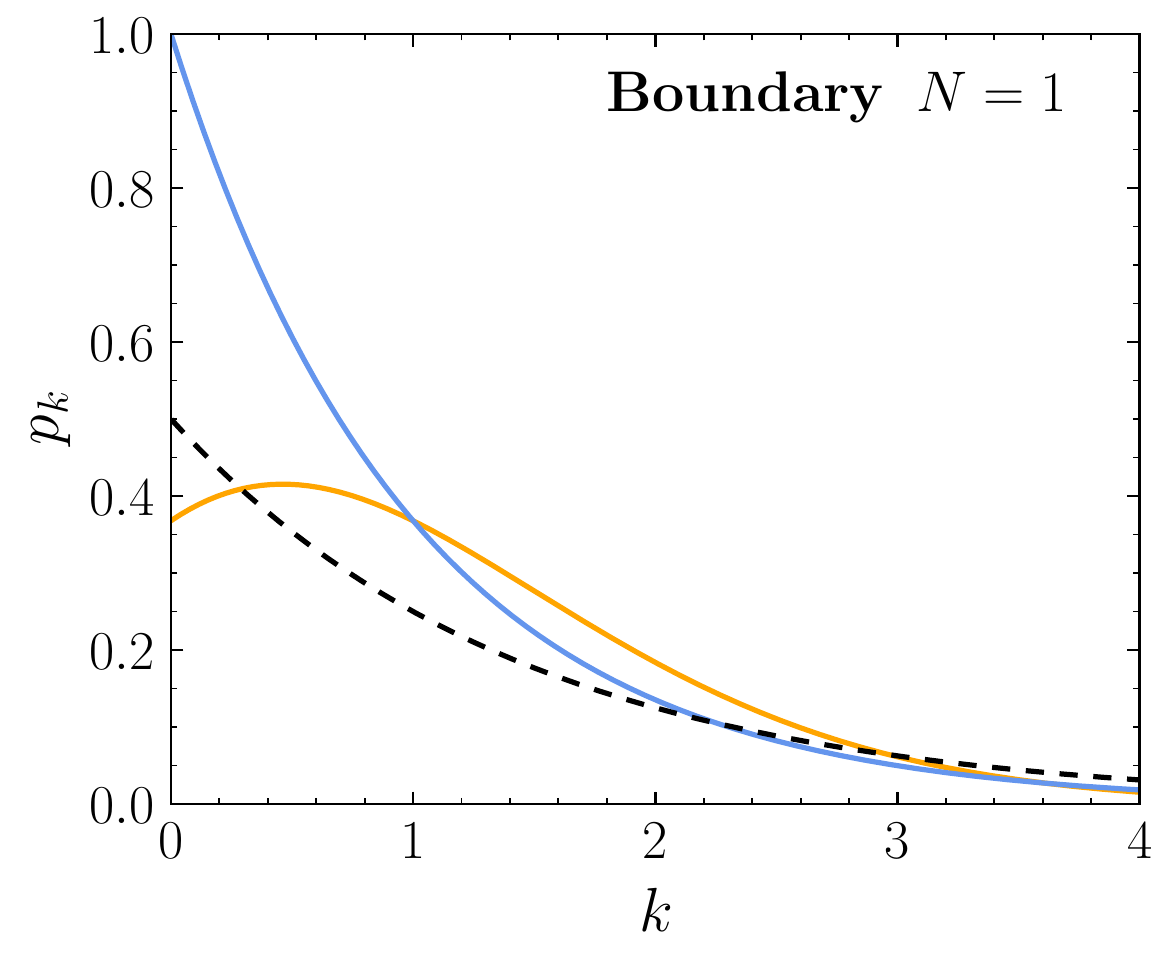}
\hspace{0.3cm}
\includegraphics[scale=0.28]{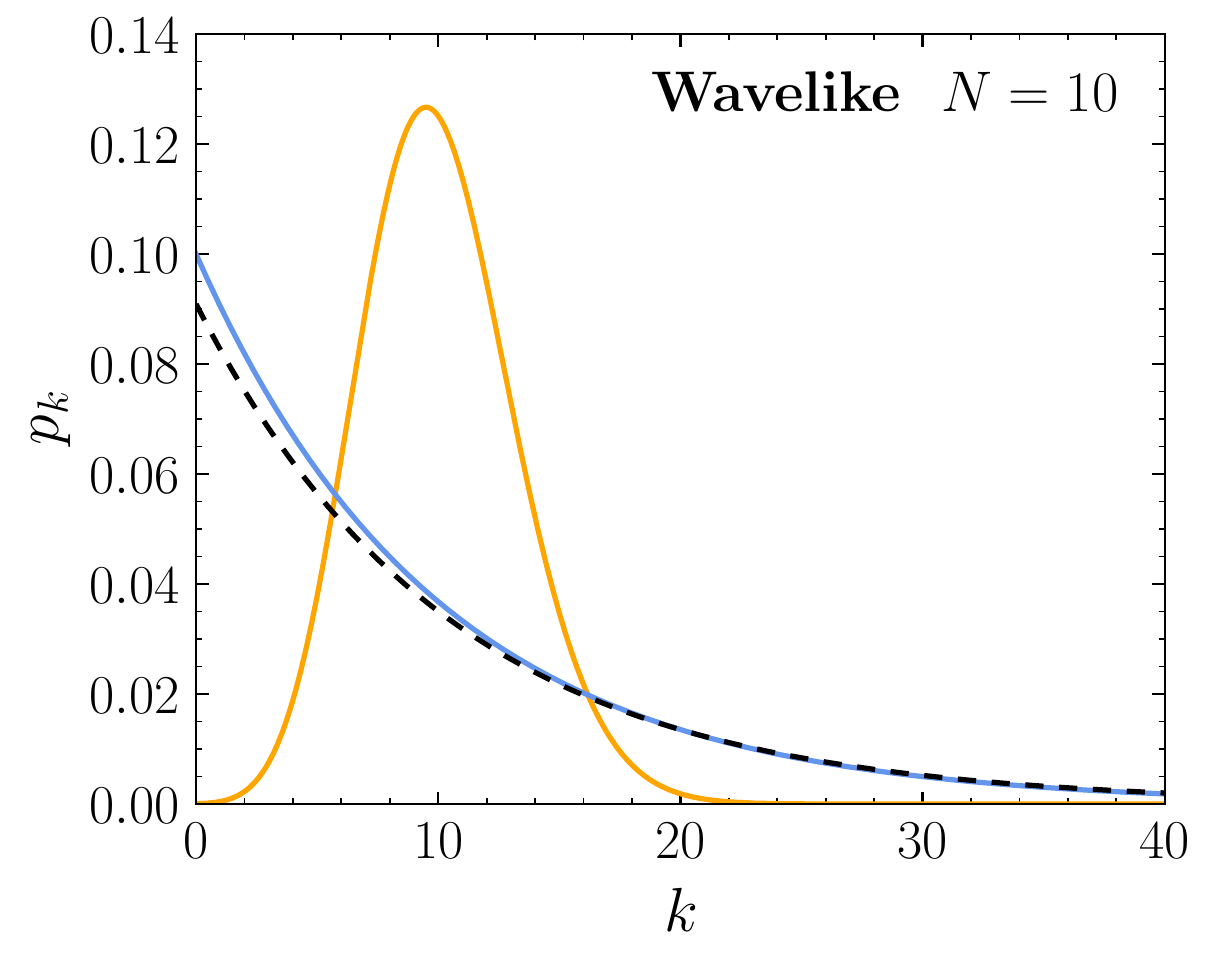}
\end{center}
\vspace{-0.7cm}
\caption{The probability of observing $k$ quanta within $V_c$ for a Gaussian $P(\alpha)$ (black dashed), Poisson (orange), or exponential (blue) distribution.
Distributions are analytically continued to an arbitrary real $k \geq 0$.
We show results for three different values of the mean: 0.1 (left), 1 (middle), and 10 (right).
In the particle regime ($N \ll 1$) the Gaussian density matrix is well described by a Poisson distribution, whereas in the wave regime ($N \gg 1$) the statistics become exponentially distributed.
For $N \simeq 1$ all three distributions are distinct.
}
\label{fig:WP-lims}
\end{figure*}

Although we can compute the mass at which $N=1$, we emphasize that \textit{there is no hard boundary between the wave and particle descriptions}.
Instead, as the calculations in the present section demonstrate, there is a continuous description of DM across the boundary; the expected behaviors emerge for $N \gg 1$ and $N \ll 1$, with a unique description appearing when $N \sim 1$.
Further, as emphasized in Sec.~\ref{sec:classical}, once we have a form for the density matrix, we can always perform the completely quantum calculation that includes all these limits, as exemplified by Eq.~\eqref{eq:hatphi2n}.
Again, the classical wave limit is simply a convenient approximation that holds for $N \gg 1$ and $P(\alpha) \geq 0$.

In order to demonstrate these claims, we consider the following question: how much energy is in the DM field within a given volume $V$?\footnote{This physical volume should not be confused with the volume introduced to discretize the mode expansion in Eq.~\eqref{eq:phi-finiteV}.
Further, similar to Eq.~\eqref{eq:hatphi2n}, we imagine the system has been regulated so there is no zero-point contribution to the energy; see App.~\ref{app:zeropoint}.}
The obvious answer is simply $\rho V$.
However, as usual it is the fluctuations rather than the mean that encode the interesting information.
In order to study the fluctuations, we focus on a single mode of the field, which we take to have energy $m$.
A single mode proves sufficient to understand the general behavior of DM across the boundary, although we show the impact of the full set of modes in a detailed calculation presented in App.~\ref{app:WP-Full}.
In the case of a single mode, the amount of energy in the field is equivalent to asking how many DM quanta $k$ appear in the region, with the energy then simply being $mk$.
Accordingly, we would heuristically expect that in the particle regime ($N \ll 1$) the number of quanta and hence energy should be Poisson distributed, as it is the result of a counting experiment for the number of particles in the volume.
In the wavelike regime ($N \gg 1$), we have already computed in Eq.~\eqref{eq:rho-stats} that the energy density should be exponentially distributed.
We now confirm that both of these expectations emerge continuously from the full quantum description when we study the system in a volume comparable to its coherence volume.
We further demonstrate two additional points: 1. For $N \simeq 1$ the fluctuations are neither exponential nor Poisson; and 2. For $V \gg V_c$ the fluctuations become Gaussian, however with a variance that varies dramatically between the wave and particle limits.

At the outset, we emphasize that we are imagining performing the measurement of the energy in a thought experiment rather than considering $V$ as an actual detector volume, in the wavelike regime this question has been considered in searches for DM with interferometers~\cite{Kim:2023pkx}, pulsar timing arrays~\cite{Kim:2023kyy}, and astrometry~\cite{Dror:2024con,Kim:2024xcr}.
If we were considering the measurements in an actual detector, accounting for the weak coupling between DM and the detector is crucial, as it determines how the fluctuations in the DM field are translated into observable fluctuations in our detector, see e.g. Ref.~\cite{Carney:2023nzz}.

Returning to the question of interest, we assume that the density matrix of our single mode takes the Gaussian form of Eq.~\eqref{eq:GaussianRho} (deviations from this are discussed in the next section).
Introducing a pair of complete Fock states, we can rewrite the Gaussian density matrix in the number basis as,\footnote{When extended to multiple modes, Eq.~\eqref{eq:rho-Fockbasis} is the density matrix that was assumed in Ref.~\cite{Blinov:2024jiz}.
In that work it was speculated that this may be an appropriate density matrix for DM, which we see from the present work is equivalent to the assumption that $P(\alpha)$ is Gaussian.}
\be
\hat{\rho} = \frac{1}{1+N} \sum_{k=0}^{\infty} \left(\frac{N}{1+N} \right)^{k} \vt k \ra \la k \vt.
\label{eq:rho-Fockbasis}
\ee
Importantly, the density matrix is diagonal in the Fock basis.
We emphasize this is not generic and it is certainly not true for a pure coherent state, see App.~\ref{app:Pa-Evo} for further discussion.
Regardless, it implies that we can immediately read off the probability of observing $k$ quanta,
\be
p_k = \frac{N^k}{(1+N)^{k+1}}.
\label{eq:pk-Vc}
\ee
From here we can directly infer the quanta and energy statistics.
However, this $p_k$ does not describe the statistics in an arbitrary volume, it describes them within the coherence volume $V_c$; indeed, as discussed above, $N$ is the number of states in the coherence volume and the above distribution obeys $\la k \ra = N$.
Accordingly, Eq.~\eqref{eq:pk-Vc} can be used to study fluctuations within $V_c$, which is sufficient to observe an interesting transition across $N=1$.
We extend the discussion to a more general volume and justify the above association shortly.

The mean of Eq.~\eqref{eq:pk-Vc} is $\mu_k = N$, exactly as expected from the mean energy density.
The interesting behavior resides in the fluctuations.
With this in mind, we turn to the variance in the number of quanta, which is here $\sigma_k^2 = N(N+1)$.
For $N \ll 1$, $\sigma_k^2 \simeq N = \mu_k$ whereas for $N \gg 1$, $\sigma_k^2 \simeq N^2 = \mu_k^2$.
As expected, these are the variances of the Poisson and exponential distributions, respectively, as can be confirmed from the appropriate distributions,
\be
p_{k,P} = \frac{N^k e^{-N}}{k!},
\hspace{0.5cm}
p_{k,E} = \frac{1}{N} e^{-k/N},
\ee
where $k$ is a continuous real variable for the exponential distribution.
For $N\simeq 1$ neither distribution is appropriate and instead the full expression in Eq.~\eqref{eq:pk-Vc} is required.
This is highlighted in Fig.~\ref{fig:WP-lims}, where we show the three distributions (analytically continued to arbitrary real $k \geq 0$).

The above results are suggestive that the full distribution becomes Poisson or exponential in the particle and wavelike limits, but they do not formally demonstrate that the distributions match.
In particular, the higher moments may not agree.
To study this carefully, it is convenient to introduce the moment generating function (MGF), $M(t) = \la e^{t k} \ra$ for $t \in \mathbb{R}$.
From the definition we can extract the moments through $\la k^n \ra = \partial_t^n M(t) \vt_{t=0}$, with $M(0)=1$ being equivalent to the normalization of the probability distribution.
For the MGF to exist we require $\la e^{t k} \ra > 0$ to exist in an open region around $t=0$.
Turning to our explicit distribution for the Gaussian density matrix, Eq.~\eqref{eq:pk-Vc} implies that the generating function for the number of quanta in the coherence volume is,
\be
M^c_G(t) = \frac{1}{1+N(1-e^t)},
\ee
where $t < \ln(1+1/N)$ to ensure $M^c_G(t) > 0$.

The MGF satisfies a number of important properties.
One that we can exploit immediately is that if the MGF of two distributions is equal, then the distributions themselves must be equivalent.
With this in mind, note that the MGF of the Poisson and exponential distributions of mean $N$ are given by
\be
M_P(t) = e^{N(e^t-1)},
\hspace{0.5cm}
M_E(t) = \frac{1}{1-Nt},
\label{eq:MPE}
\ee
where there is no restriction on $t$ for $M_P(t)$, whereas for $M_E(t)$ we require $t < 1/N$.
We can use these results to establish a formal equality between the distributions in the wave and particle regime.
In the particle limit, we have
\be
\lim_{N \ll 1} M^c_G(t) \simeq \lim_{N \ll 1} M_P(t) \simeq 1+N(e^t-1),
\ee
establishing that the Gaussian becomes exactly Poisson.
(Note that as $N \to 0$ the restriction on $t$ for $M^c_G(t)$ is removed.)
In the wavelike regime the limit must be taken more carefully.
For instance, naively taking $N \to \infty$ at fixed $t$ in either $M^c_G(t)$ or $M_E(t)$ leads to a result that fails even the basic normalization condition of $M(0)=1$.
The point that is missed is that in all evaluations of the MGF we eventually take $t \to 0$ which can compensate for a large $N$.
Indeed, in both cases we see that as $N \to \infty$, the restriction on $t$ leaves less and less room for an open neighborhood around the origin.
Therefore, large $N$ forces small $t$, so that,
\be
\lim_{N \gg 1} M^c_G(t) \simeq \lim_{N \gg 1} M_E(t) = \frac{1}{1-Nt}.
\ee
This establishes the connections suggested in Fig.~\ref{fig:WP-lims}, although we emphasize that formal equality holds only in the limit $N \to 0$ or $N \to \infty$.
The general distribution is neither Poisson nor exponential.

\begin{figure}[!t]
\begin{center}
\includegraphics[scale=0.42]{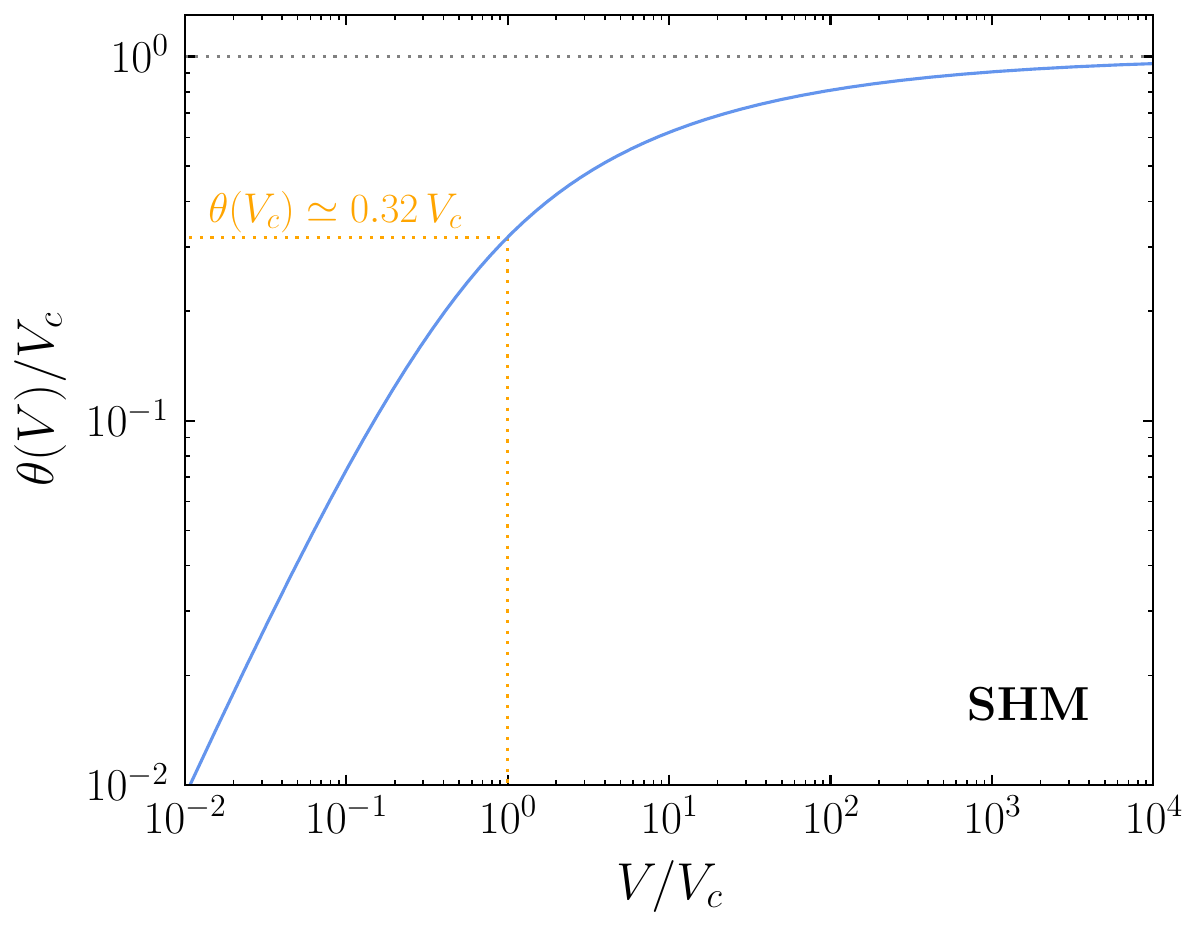}
\end{center}
\vspace{-0.7cm}
\caption{The form of $\theta(V)$ as introduced in Eq.~\eqref{eq:musig-generalV} for the SHM.
This function, defined in App.~\ref{app:WP-Full}, controls the behavior of the system as we study it at volumes across the coherence volume, $V_c$.
The asymptotic behavior of the function, as given in Eq.~\eqref{eq:thetaV-limits}, is independent of the properties of the scalar field, whereas the exact transition for $V$ across $V_c$ does depend on the properties of the field.}
\label{fig:Theta-V}
\end{figure}

Let us now extend the discussion to a more general volume.
Using the formalism developed so far a fully quantum calculation in an arbitrary volume can be performed.
The calculation is slightly extended, so we defer it to App.~\ref{app:WP-Full}, although as shown there the mean and standard deviation for the number quanta is,
\be
\mu_k = \bar{n} V,\hspace{0.5cm}
\sigma_k^2 = \bar{n} V\,[ \bar{n}\, \theta(V) + 1 ].
\label{eq:musig-generalV}
\ee
Here $\theta(V)$ is a function of $V$ that has units of volume.
An exact definition of $\theta(V)$ is provided in App.~\ref{app:WP-Full}; in general it depends on the form of $g^{(1)}(\tau,\bd)$, although for the SHM we can compute it exactly, and find the form shown in Fig.~\ref{fig:Theta-V}.
Broadly, we can summarize the function as,
\be
\theta(V) \simeq \left\{ \begin{array}{lc} V & V \ll V_c, \\ 0.32 V_c & V \sim V_c, \\ V_c & V \gg V_c. \end{array}\right.
\label{eq:thetaV-limits}
\ee
The asymptotic behavior of $\theta(V)$ is independent of $g^{(1)}$, although the exact transition between these limits does depend on the exact form.

Using this more general result, we can confirm that for $V = V_c$, we have $\mu_k = \bar{n} V_c = N$ and $\sigma_k^2 \simeq N(0.32 N+1)$, which agrees with our earlier heuristic discussion up to an ${\cal O}(1)$ correction.
We can now consider other volumes, however.
Firstly, if we assume $V \ll V_c$, then using Eq.~\eqref{eq:thetaV-limits} the results in Eq.~\eqref{eq:musig-generalV} become,
\be
V \ll V_c:\hspace{0.2cm}
\mu_k = \bar{n} V = N N_c,\hspace{0.5cm}
\sigma_k^2 = \mu_k\,( \mu_k + 1 ).
\ee
Here, we introduced the number of coherence volumes, $N_c = V/V_c$, in order to rewrite the mean in terms of $N$.
From this we see that if $\mu_k = N N_c \gg 1$, the statistics are exponentially distributed, whereas for $N N_c \ll 1$ the variance reduces to the Poisson limit.
For the limit under consideration, $N_c \ll 1$, and so we could only have exponential behavior if $N \gg 1/N_c \gg 1$.
Equivalently, exponential statistics can only occur for DM sufficiently in the wavelike regime, although even then as $V$ shrinks eventually the fluctuations become Poisson.

Next, consider the case where $V \gg V_c$.
As in the small volume limit, we expect there must be a deviation from the exponential statistics as the volume increases.
Roughly, from Eq.~\eqref{eq:dDM-SHM} 1\,$\mu$eV DM should exhibit wavelike phenomena at a characteristic scale of $\sim$500\,m, however, if we study the system at Galactic scales, there should be a deviation in this behavior as it is no longer possible to build up coherent fluctuations over the full volume.
The transition in behavior as a function of volume is smooth, as Fig.~\ref{fig:Theta-V} shows, although asymptotically we can treat the field as being a combination of $N_c$ statistically independent volumes.
More quantitatively, from the general results above,
\be
V \gg V_c:\hspace{0.2cm}
\mu_k = N N_c,\hspace{0.5cm}
\sigma_k^2 = \mu_k\,[ N + 1 ].
\label{eq:musig-VggVc}
\ee
For particle-like DM, with $N \ll 1$, the fluctuations have a Poisson variance, $\sigma_k^2 = \mu_k$.
In the wavelike regime, $N \gg 1$, the variance is neither Poisson nor exponential, with $\sigma_k^2 = N_c N^2 < \mu_k^2$; in fact, as we show below they are Erlang or approximately normally distributed.

To formalize the study of the system when $V \gg V_c$ we utilize an additional feature of the MGF: the generating function of the sum of a set of independent random variables is given by the products of their individuals MGFs.
We can exploit this by treating the system as $N_c$ individual coherence volumes.
In each of these volumes, the number of quanta is independent and described by Eq.~\eqref{eq:pk-Vc}.
Accordingly, the MGF for the number of quanta in a volume $V$ is
\be
M^V_G(t) = [M_G^c(t)]^{N_c} = \left[\frac{1}{1+N(1-e^t)} \right]^{N_c}\!,
\ee
again with $t < \ln(1+1/N)$.
The associated probability distribution can be determined as,
\be
p_k = \frac{(k+N_c-1)!}{k!(N_c-1)!} \frac{N^k}{(1+N)^{k+N_c}}.
\label{eq:pk-genNc}
\ee
From either the MGF or probability distribution we obtain a mean and variance of $\mu_k = N_c N$ and $\sigma_k^2 = N_c N(N+1)$, in exact agreement with Eq.~\eqref{eq:musig-VggVc}, validating our treatment of each coherence volume as independent.
The above analysis emphasizes why the statistics remain Poisson in the particle regime, as a sum of Poisson distributions is itself Poisson distributed.

Consider the limiting behavior of the statistics in a large volume, $V \gg V_c$.
For particle DM, we wish to take $N$ small and $N_c$ large.
In order to keep the mean fixed, we take $N \to \epsilon N$ and $N_c \to N_c/\epsilon$, so that taking $\epsilon \to 0$ sends $N_c \to \infty$ whilst leaving $\mu_k = N_c N$ constant.
Doing so, we obtain,
\be
\lim_{\epsilon \to 0} \left[\frac{1}{1+\epsilon N(1-e^t)} \right]^{N_c/\epsilon} = e^{N_c N(e^t-1)}.
\ee
Comparing with Eq.~\eqref{eq:MPE}, we see that the distribution has become exactly Poisson with mean $N_c N$.
In the wavelike limit, as before, we effectively restrict $t$ to a narrower range, which implies,
\be
\lim_{N \gg 1} M^V_G(t) \simeq \left[\frac{1}{1-Nt} \right]^{N_c} = [M_E(t)]^{N_c}\!.
\ee
That this is the product of $N_c$ exponential generating functions of course had to be the case, but as written we can also recognize this as the MGF of the Erlang distribution, which describes a sum of $N_c$ exponential distributions.

Formally, the above resolves what happens as we combine $N_c$ coherence volumes: in the particle regime we have a sum of Poisson distributions, which is itself Poisson, whereas in the wavelike paradigm we have a sum of exponentials, which is Erlang distributed.
What happens as we take $N_c$ to be larger and larger?
As both the Poisson and exponential distributions obey the central limit theorem, as we add more and more coherence volumes, both distributions must tend to a  normal distribution.
The normal distributions are not identical, however.
In fact, we can apply the central limit theorem to a sum of $N_c$ draws from the Gaussian $P(\alpha)$ prediction given in Eq.~\eqref{eq:pk-Vc}.
Doing so, we arrive at a normal distribution with mean $\mu_k = N_c N$ and variance $\sigma_k^2 = N_c N(N+1)$.

Accordingly, for $V \gg V_c$ the number of quanta (and hence energy) in the volume undergoes Gaussian fluctuations regardless of the value of $N$.
But the size of those fluctuations encodes the nature of DM, as would be revealed by a measurement of the variance over the mean, $\sigma_k^2/\mu_k = N+1$.
Equivalently, consider the ratio $\sigma_k/\mu_k =\sqrt{(N+1)/\mu_k}$.
In the particle limit, fluctuations exhibit the conventional Poisson suppression of $\sigma_k/\mu_k \sim 1/\sqrt{\mu_k}$, whereas in the wavelike limit $\sigma_k/\mu_k \sim 1/\sqrt{N_c} \gg 1/\sqrt{\mu_k}$.
In short, in the wavelike limit the fluctuations are much larger than expected from Poisson distributed particle DM, although less than expected of an exponential distribution, which is recovered when $V = V_c$.
Effectively, for wavelike DM, the fact that separate coherence volumes are incoherent prevents even larger fluctuations being generated, although the fact there were significant fluctuations within each $V_c$ leaves a fingerprint on large scales.
For $m \ll 18.7\,\eV$, the variance is enhanced beyond that the particle-like Poisson case by a factor of $N+1 \simeq 10^{29} (1\,\mu\eV/m)^4$, from Eq.~\eqref{eq:N}.
Of course, even if there is a significantly larger fluctuation than expected, they only persist for the coherence time given in Eq.~\eqref{eq:tDM-Q}, which can be short ($\tau_{\DM} \ll 1\,\textrm{s}$) even when $N \gg 1$.

A related question to the one explored above is how many DM quanta would we expect to measure at a perfectly efficient detector in a fixed time interval $T$.
This question is the direct time analog of the spatial study above; we expect the system to behave coherently up to $T \simeq \tau_c$, and then act independently between these intervals.
Indeed, as reviewed in Ref.~\cite{Mandel:1995seg}, the behavior in the time domain is exactly the same to that derived above.
The probability of observing $k$ quanta is given by Eq.~\eqref{eq:pk-genNc} with $N_c \simeq \textrm{min}[1,T/\tau_c]$ and there is an equivalent function to $\theta(V)$ which allows a smooth transition between the regimes.
Once corrected for finite detector efficiency, these results can be used to determine the exact pattern of fluctuations expected for detectors counting discrete DM induced events, even right at the wave-particle boundary as studied in Ref.~\cite{Arvanitaki:2017nhi}.

In summary, in this section we have performed an explicit calculation across the wave-particle boundary, demonstrating that much of the intuitive behavior we would expect holds, and also that near the $N \simeq 1$ boundary the behavior of DM is unique.
Of course, the calculations we considered, namely the energy in a box or the number of DM quanta counted in a given time, are rather contrived.
Further, such measurements can only be rendered by an experiment.
A more complete discussion would necessitate coupling the DM to a detector and drawing on techniques from quantum measurement theory (for a discussion of this in the context of DM, see e.g. Ref.~\cite{Beckey:2023shi}).
This is an interesting direction to pursue, but the results of this section already establish that in principle there is no obstacle to computing the properties of DM for an arbitrary mass and hence $N$, without resorting to an assumption that it behaves as a wave or particle.

%%%%%%%%%%%%%%%%%%%%%%%%%%%%%%%
\section{Non-Gaussianities and other forms of the density matrix}
\label{sec:NG}
%%%%%%%%%%%%%%%%%%%%%%%%%%%%%%%

So far we have primarily focused on the implications of DM being described by the density matrix with a Gaussian $P(\alpha)$ given in Eq.~\eqref{eq:GaussianRho}.
There are a number of reasons to find this form attractive.
Regardless of the state DM was born in the system has undergone considerable evolution.
Most significantly, the process of virialization is a violent one~\cite{Lynden-Bell:1966zjv}.
Even if $P(\alpha)$ for DM was non-Gaussian before the galaxy formed, through formation the DM field could have been fragmented and randomized, in which case the local DM field at the present epoch could be treated as a large sum of independent fields, suggesting the quantum central limit theorem could apply.
Further, if the DM ever thermalized, then as we review in App.~\ref{app:Pa-Evo}, there are examples where the evolution to a Gaussian density matrix can be explicitly computed.

Beyond the above suggestive arguments, we offer no proof that the DM density matrix takes a Gaussian form.
Resolving this represents an important step in determining the behavior of wave DM and in confirming -- or refuting -- our various claims.
A path to doing so would be to study the cosmological evolution of the density matrix for various assumptions of its initial form.
There is a wide literature on the topic, delving into the theories of open quantum systems, master equations, and gravitational decoherence~\cite{Allali:2020ttz, Allali:2020shm, Allali:2021puy,Kopp:2021ltb,Cao:2022bua,Cao:2022kjn,Eberhardt:2023axk}.

Although we leave aside a detailed study of the evolution of the DM density matrix, one brief comment we would make is that studying the evolution could also represent an opportunity to understand the emergence of non-trivial states in wave DM, such as solitons and large scale Bose-Einstein condensates (BEC), see e.g. Refs.~\cite{Semikoz:1995rd,Sikivie:2009qn,Erken:2011dz,Guth:2014hsa}.
The only point we have to mention here is that if the system is ever driven into a condensate where a single mode dominates, from the discussion in Sec.~\ref{sec:coherence} we would expect that the autocorrelation function would tend to a constant.
This implies that if we live within such a system -- for instance, it has been proposed there could be a large soliton around the Sun, see e.g. Refs.~\cite{Banerjee:2019epw,Budker:2023sex} -- the correlation time and length of DM could be significantly larger than would be inferred from the SHM as given in Eqs.~\eqref{eq:tDM-Q} and \eqref{eq:dDM-SHM}, which generically leads to a signal that is more easily detected.
If the BEC were further described by a Gaussian $P(\alpha)$, then following the discussion from the previous section we would expect large exponential fluctuations in the density (or equivalently the number of quanta observed in a given time) over the now enhanced coherence scales.

The more general point we wish to emphasize in this section is that the assumed form of $P(\alpha)$ impacts experimental measurements.
Specifically, following the discovery of DM, one could look to constrain $P(\alpha)$ and hence $\hat{\rho}$. 
We do not attempt to quantify to what extent one could constraint the DM density matrix in the present work, instead our focus is simply to outline how a variation to $P(\alpha)$ can imprint itself on experimental measurements (see also Ref.~\cite{Marsh:2022gnf}).
In particular, a non-Gaussian $P(\alpha)$ would induce non-Gaussianities in various DM observables in the form of deviations from the Gaussian predictions we have determined.
If the density matrix has not evolved to a Gaussian form, it may retain information as to the state it was born in, which could then be accessed experimentally.
As exciting as this would be, we emphasize once more that our expectation is that $P(\alpha)$ is Gaussian.
Nevertheless, to highlight how the deviations from Gaussian behavior could emerge we consider several examples below.

Consider first the example studied in Sec.~\ref{sec:boundary} where we counted the number of quanta in a given volume.
The results there followed from the number of quanta the Gaussian $P(\alpha)$ predicts within the coherence volume, which is given by Eq.~\eqref{eq:pk-Vc}.
As an extreme example, if instead the DM is in a Fock state of $N$ quanta, there would be no variance observed in the fluctuations in $V_c$ or a larger volume.
Alternatively, if the system is in a pure coherent state, $\hat{\rho} = \vt \alpha \ra \la \alpha \vt$, the probability to observe $k$ quanta in $V_c$ is given by a Poisson distribution,
\be
p_k = \frac{|\alpha|^{2k} e^{-|\alpha|^2}}{k!},
\ee
which holds independently of the wave-particle boundary.
As the sum of Poisson distributions is Poisson, the number of quanta and the energy in the volume are Poisson distributed over any scale.
These represent clear deviations from the Gaussian predictions for DM in the wavelike regime discussed in the previous section.

Quadrature rather than number measurements of the DM field would also differ were DM described by a pure coherent state.
As discussed around Eq.~\eqref{eq:XY-quad}, in the Gaussian case each quadrature is itself a normally distributed variable with zero mean.
For a pure coherent state in the wavelike limit, we have (for a single mode) $\hat{X} \simeq \sqrt{2} \operatorname{Re}[\alpha]$ and $\hat{Y} \simeq \sqrt{2} \operatorname{Im}[\alpha]$, but now $\alpha$ is a fixed rather than stochastic variable, so that repeated measurements of the field quadratures do not fluctuate.
Further examples can be obtained by interfering systems with different distributions.
If the system results from the superposition of two independent fields described by $P_1(\alpha)$ and $P_2(\alpha)$ then the total density matrix is described by~\cite{Mandel:1995seg}
\be
P(\alpha) = \int d\alpha'\,P_1(\alpha') P_2(\alpha-\alpha').
\label{eq:2convolution}
\ee
For example, the superposition of two independent fields described by coherent states $\vt \beta_1 \ra$ and $\vt \beta_2 \ra$ is again described by a coherent state $\vt \beta_1+\beta_2 \ra$.
If we instead combine a system described by a coherent state $\vt \alpha_0 \ra$ with one determined by the Gaussian $P(\alpha)$, the total density matrix is described by
\be
P(\alpha) = \frac{1}{\pi N} e^{-|\alpha-\alpha_0|^2/N}.
\ee
As before this Gaussian has variance $N/2$, but unlike in Eq.~\eqref{eq:GaussianRho}, the value of $\alpha$ now has a non-zero mean.
If such a system described DM we would obtain Gaussian fluctuations of the field, but now $\la \phi \ra \neq 0$ and similarly for its derivatives.

A remarkable possibility would be if DM had a $P(\alpha)$ that is explicitly quantum and therefore not positive definite.
As shown in Sec.~\ref{sec:classical}, this would imply that DM cannot be described as a classical random field even if it has large occupation as there are measurement outcomes such a state can generate that no classical field could reproduce.
In practice, however, experimentally measuring any such behavior would be extremely challenging.
The reason for this is the experimental measurements needed to confirm a negative $P(\alpha)$ are generically of the form of measuring a smaller than classical variance in a given observable.
An example would be if the system is in a Fock state, the number of quanta observed does not fluctuate, or measuring $g^{(2)}(0,\mathbf{0}) < 1$ as discussed in Sec.~\ref{sec:g1g2}.
A considerable complication to any such observation, however, is to include and carefully account for the quantum mechanics of the measurement device.
Generically, the measurement device has its own series of fluctuations that can swamp the statistics of the field being measured, especially in the case where that field only couples weakly to the apparatus.
For example, it was shown in Ref.~\cite{Carney:2023nzz} that such considerations present an enormous barrier to any attempt to infer a quantum nature of gravity through graviton number or quadrature measurements.
Although we hope DM couples to us far more strongly than gravity, the general points raised in that work suggest it would be equally challenging to measure any signal of a negative $P(\alpha)$ for DM.
As such, although we do not attempt to prove it here, we strongly suspect that even if DM were described by a quantum $P(\alpha)$ there must remain an appropriate classical wave description of the field that is experimentally indistinguishable for all intents and purposes.

%%%%%%%%%%%%%%%%%%%%%%%%%%%%%%%
\section{Discussion}
\label{sec:discussion}
%%%%%%%%%%%%%%%%%%%%%%%%%%%%%%%

The discovery of wave DM could be imminent, raising the importance of understanding its behavior in detail and going beyond the commonly adopted ansatz of $\phi(t) \sim \cos (m t)$.
In this work we have outlined a procedure for doing so where DM is described quantum mechanically.
Starting from a density matrix, which we argued likely takes a Gaussian form, we can perform calculations at arbitrary values of the DM mass without ever assuming it behaves as a classical wave.
Yet the picture also clarifies how when $m \ll 10\,\eV$ DM can be described as a classical stochastic field, with its fluctuations inherited from the underlying density matrix.
Further, drawing on the quantum optics literature we can formalize the various discussions of the coherence properties of the classical wave picture of DM, including the coherence time, length, and volume.

Ultimately, however, these are only early steps towards formalizing the description of wave DM.
We therefore end with a brief discussion of the open questions suggested by the present work.
The most pressing is to resolve the two clear shortcomings of our analysis.
The first of these is to study how the various properties of DM that are clarified with a quantum mechanical description are actually imprinted onto a detector.
For this one can use techniques from quantum measurement theory to simultaneously account for the quantum mechanics of DM and the detector (see e.g. Ref.~\cite{Beckey:2023shi} and also Ref.~\cite{Bernal:2024hcc}).
For instance, it would be interesting to determine how the large fluctuations in the observed number of DM particles studied in Sec.~\ref{sec:boundary} could imprint themselves on a detector that measures individual quanta in search of wavelike DM.

Secondly, it will be important to formally establish what the density matrix of DM actually is.
We have argued that $\hat{\rho}$ likely takes a Gaussian form and have further noted in Sec.~\ref{sec:NG} that deviations from this would imprint themselves as non-Gaussian fluctuations in DM measurements.
Nevertheless, formally one could look to evolve the density matrix of DM from the early Universe until today, which would resolve whether there remains any hint of the state that DM was born in.
(In App.~\ref{app:Pa-Evo} we discuss how the evolution proceeds in a particularly simple example.)
Such an analysis could further prove useful in understanding the evolution of DM into configurations such as a BEC.

Lastly, one could imagine extending the techniques discussed in this work beyond scalar DM.
An obvious extension is to dark-photon DM, where most of what we have discussed would apply almost immediately.
The only obstacle is in determining the appropriate distribution for the polarization of the field, a question around which there is presently considerable uncertainty, see e.g. Ref.~\cite{Caputo:2021eaa}, and one could hope that quantum optics techniques could prove useful.
The formalism could also be extended to study relativistic bosonic states, such as axions produced in the early Universe or gravitational waves radiated from black holes.
With this in mind, where possible we have stated results in a manner that did not assume the field is non-relativistic, and indeed in certain cases the form of $P(\alpha)$ is clear; for example, axions thermally emitted from the Sun should have a Gaussian density matrix.
More speculatively, an extension to fermions may be possible, which could open an alternative path to studying questions such as the coherence properties of neutrinos.
Of course, when turning to fermions the approach in this paper would be modified at step one as the coherent state is an inherently bosonic object.
Nevertheless, Cahill and Glauber showed in Ref.~\cite{PhysRevA.59.1538} that a similar description applies for a fermionic field, although it has not been as developed as the bosonic analog.

%%%%%%%%%%%%%%%%%%%%%%%%%%%%%%%
\section*{Acknowledgements}
%%%%%%%%%%%%%%%%%%%%%%%%%%%%%%%

We thank Cliff Burgess, Dan Carney, Joshua Combes, Hyungjin Kim, Shirley Li, Nadav Outmezguine, Gilad Perez, Ryan Plestid, and Gray Rybka for useful discussions.
We further thank the organizers of the 2022 CERN-CKC Theory workshop on Jeju Island where this work was initiated.
The work of DYC was supported by the National Research Foundation of Korea (NRF) grant funded by the Korea government (MSIT) (RS-2024-00340153), and the CERN-Korea Theoretical Physics Collaboration and Developing Young High-Energy Theorists fellowship program (NRF-2012K1A3A2A0105178151).
The work of NLR was supported by the Office of High Energy Physics of the U.S. Department of Energy under contract DE-AC02-05CH11231.
LTW is supported by the Department of Energy grant DE-SC0013642.
Part of this work was performed at Aspen Center for Physics, which is supported by National Science Foundation grant PHY-2210452.

\appendix
%%%%%%%%%%%%%%%%%%%%%%%%%%%%%%%
\section{The Quantum Central Limit Theorem}
\label{app:Quantum-CLT}
%%%%%%%%%%%%%%%%%%%%%%%%%%%%%%%

In this appendix we review the classic argument of Glauber~\cite{Glauber:1963tx} that $P(\alpha)$ obeys an analog of the central limit theorem.
This argument underpinned the Gaussian form of the density matrix in Eq.~\eqref{eq:GaussianRho} that we argued likely holds for DM and therefore used extensively through the main text.
In particular, we imagine that the local DM field is a superposition of a large number $n$ of independent and identically distributed random fields.
If DM has a thermal origin this condition is satisfied, but beyond this we offer only heuristic arguments that this is true as discussed in Sec.~\ref{sec:NG}.
If we assume that the contribution of each field to the DM is described by a Glauber-Sudarshan distribution $\pi(\alpha)$, and further following Ref.~\cite{Glauber:1963tx} assume the strong stationarity condition of $\pi(\alpha) = \pi(|\alpha|)$, then in a manner analogous to the proof of the classical central limit theorem, we establish that $P(\alpha)$ must be a Gaussian as $n \to \infty$.

To begin with, by a direct generalization of Eq.~\eqref{eq:2convolution}, the quasi-probability distribution that describes a field that is the superposition of $n$ identical and independent fields is given by convolving the individual distribution $n$ times.
In detail,
\be
P(\alpha) = \int \delta\Big(\alpha - \sum_{j=1}^n \alpha_j \Big) \prod_{j=1}^n d\alpha_j\, \pi(\alpha_j).
\label{eq:CLT-step1}
\ee

Taking the Fourier transform of this result, we can replace the convolution by a product.
Before doing so, let us introduce a convenient, albeit unorthodox, Fourier transform convention for the present discussion.
Using subscripts $r$ and $i$ to refer to the real and imaginary parts of complex variables, we define the transformation between a quasi-probability distribution $g(\alpha)$ and its transform $\tilde{g}(\lambda)$ by,\footnote{We caution that a different Fourier transform convention is used in the main text (cf. Eq.~\eqref{eq:PSD-def}) and in App.~\ref{app:wigner}.
The choice here is made so that $\tilde{g}(\lambda)$ is comparable to the characteristic function of a classical probability distribution.}
\bea
\tilde{g}(\lambda) &= \int d\alpha\,e^{i (\alpha_r \lambda_r+\alpha_i \lambda_i)/\sqrt{n}}\,g(\alpha), \\
g(\alpha) &= \int \frac{d\lambda}{n(2\pi)^2}\,e^{-i (\alpha_r \lambda_r+\alpha_i \lambda_i)/\sqrt{n}}\,\tilde{g}(\lambda).
\label{eq:CLT-Fourier}
\eea
The factors of $n$ are added for later convenience.
Accordingly, the Fourier transform of Eq.~\eqref{eq:CLT-step1} is,
\be
\tilde{P}(\lambda) = \left[ \tilde{\pi}(\lambda) \right]^n\!.
\label{eq:CLT-prod}
\ee

From here, we consider $\tilde{\pi}(\lambda)$.
Expanding around $\lambda=0$ we obtain,
\begin{align}
\tilde{\pi}(\lambda) &= \tilde{\pi}(0) + \lambda_r \tilde{\pi}^{(1,0)}(0) + \lambda_i \tilde{\pi}^{(0,1)}(0) \\
&+ \frac{1}{2} \lambda_r^2 \tilde{\pi}^{(2,0)}(0)
+ \lambda_r \lambda_i \tilde{\pi}^{(1,1)}(0)
+ \frac{1}{2} \lambda_i^2 \tilde{\pi}^{(0,2)}(0) + \ldots \nonumber
\label{eq:CLT-Taylor}
\end{align}
with $\tilde{\pi}^{(p,q)}(\lambda) = \partial^p_{\lambda_r} \partial^q_{\lambda_i} \tilde{\pi}(\lambda)$.
We can relate these expressions back to $\pi(\alpha)$ using Eq.~\eqref{eq:CLT-Fourier}.
Firstly, $\tilde{\pi}(0) = \int d\alpha\, \pi(\alpha) = 1$.
Further,
\be
\tilde{\pi}^{(p,q)}(0) = \left( \frac{i}{\sqrt{n}} \right)^{p+q} \int d\alpha\,\alpha_r^p \alpha_i^q\,\pi(\alpha).
\label{eq:tp-deriv}
\ee
From here, in order to simplify the discussion we invoke the condition that $\pi(\alpha) = \pi(|\alpha|)$.
In this case, the derivatives in Eq.~\eqref{eq:tp-deriv} vanish unless $p$ or $q$ are even, and in fact the only terms we need to second order in $\lambda$ are
\be
\tilde{\pi}^{(2,0)}(0) = \tilde{\pi}^{(0,2)}(0) = - \frac{1}{2n} \int d\alpha\,|\alpha|^2\,\pi(|\alpha|),
\ee
as all others vanish.

Returning to Eq.~\eqref{eq:CLT-prod}, we have
\be
\tilde{P}(\lambda) = \left[ 1 - \frac{1}{4n}|\lambda|^2 \la |\alpha|^2 \ra + {\cal O}(1/n^2) \right]^n\!.
\ee
In the limit $n \to \infty$, the higher order terms can be neglected, and we arrive at
\be
\tilde{P}(\lambda) \simeq \exp\left[ - \frac{1}{4}|\lambda|^2 \la |\alpha|^2 \ra \right]\!,
\ee
which after applying the inverse Fourier transform becomes,
\be
P(\alpha) \simeq \frac{1}{\pi n \la |\alpha|^2 \ra} \exp \left[ - \frac{|\alpha|^2}{n \la |\alpha|^2 \ra} \right]\!.
\ee
From here, note that $\la |\alpha|^2 \ra = \la \hat{N} \ra$ is the expected number of states associated with a single source, so that $n \la \hat{N} \ra$ is the expected number of quanta for the combined field and therefore can be identified with $N$.
This completes the connection to Eq.~\eqref{eq:GaussianRho}.

Let us end with several comments.
Firstly, we can trivially repeat the argument mode by mode to justify Eq.~\eqref{eq:Gauss-multimode}.
Secondly, notice that the argument did not require $N \gg 1$ so that it need not only apply in the wavelike DM regime.
Instead, we only require $n \gg 1$, in principle we can send $n \to \infty$ while keeping $n \la \hat{N} \ra$ fixed and small.
Lastly, as noted by Glauber, if the component fields are not identical, but have comparable variances $\la \hat{N}_j \ra$, the argument proceeds identically and in the end we find $N = \sum_j \la \hat{N}_j \ra$, demonstrating that the assumption that the $n$ fields were identical can be relaxed.

%%%%%%%%%%%%%%%%%%%%%%%%%%%%%%%
\section{Continuous versus Discrete Modes}
\label{app:contdisc}
%%%%%%%%%%%%%%%%%%%%%%%%%%%%%%%

Throughout the main text we use the continuous versus discrete decomposition of the field operator into modes interchangeably as convenient.
For completeness, in this appendix we briefly outline the connection between the two choices of modes.

To discuss the mode discretization we start from the continuous definition of the scalar field operator expanded in terms of plane waves as in Eq.~\eqref{eq:phihat}, with the commutation relation given by Eq.~\eqref{eq:com-cont}.
From both results we see that the mass dimension of the creation and annihilation operators is given by $[\hat{a}^{(\dag)}]=-3/2$.
To discretize the modes, we place the system within a box of volume $V = L^3$, such that the continuous $\bk$ modes now take on values $\bk = (2\pi/L) \bn$, with $\bn = (n_x,\,n_y,\,n_z)$ and $n_{x,y,z} \in \mathbb{Z}$.
Integration over all $\bk$ modes is achieved by summing over all values of $\bn$ and the commutation amongst different modes is implemented with a Kronecker delta amongst the integers labeling the modes.
In detail,
\be
\int \frac{d^3 \bk}{(2 \pi)^3} \rightarrow \frac{1}{V} \sum_{\bk},
\hspace{0.3cm}
\delta(\bk - \bq) \rightarrow \frac{V}{(2\pi)^{3}} \delta_{\bk, \bq}.
\label{eq:dc-mapping}
\ee
The remaining volume factors are absorbed by the creation and annihilation operators through,
\be
\hat{a}_{\bk}^{(\dag)} \rightarrow  \sqrt{V} \hat{a}_{\bk}^{(\dag)},
\ee
such that these operators are dimensionless in the discrete implementation of the problem.

Applying the three replacements above allows us to move back and forth from the continuous representation of the field and commutation relations to their discrete analogs in Eq.~\eqref{eq:phi-finiteV}.
The primary utility of the discrete representation is that it simplifies the treatment of coherent states.
In particular, in the discrete representation, the coherent state of a given mode is defined by,
\be
\hat{a}_{\bq} \vt \alpha_{\bk} \ra = \delta_{\bk,\bq}\,\alpha_{\bk} \vt \alpha_{\bk} \ra.
\ee
Importantly, as in the discrete case $\hat{a}_{\bk}$ is dimensionless, so too is the coherent state eigenvalue $\alpha_{\bk} \in \mathbb{C}$.
Clearly this cannot persist in the continuous case: just as for the annihilation operator, the coherent state eigenvalue must have mass dimension $-3/2$.
In order to achieve this, one introduces the continuous displacement operator (cf. the discrete version in Eq.~\eqref{eq:displacement-discrete}),
\be
\hat{D}(\hat{a},\alpha) = \exp \left[\int \frac{d^3\bk}{(2\pi)^3}\, (\alpha_{\bk} \hat{a}_{\bk}^{\dag} - \alpha_{\bk}^* \hat{a}_{\bk} ) \right]\!.
\ee
The continuous displacement operator generates a coherent state from the vacuum (cf. Eq.~\eqref{eq:Dvac-discrete}),
\begin{align}
|\alpha \rangle &= \hat{D}(\hat{a},\alpha) |0\rangle \\
&=  \exp\left[ - \frac{1}{2} \int \frac{d^3\bk}{(2\pi)^3}\, |\alpha_{\bk}|^2 \right]
\exp \left[\int \frac{d^3\bk}{(2\pi)^3}\, \alpha_{\bk} \hat{a}_{\bk}^{\dag} \right] \vt 0 \ra. \nonumber
\end{align}
From this definition of the continuous coherent state, we can confirm that $\hat{a}_{\bk} \vt \alpha \ra = \alpha_{\bk} \vt \alpha \ra$.
Further discussion of the treatment of continuum fields can be found in e.g. Ref.~\cite{Blow:1990amy}.

In principle, the above results allow one to perform all manipulations in the continuous representation.
Such an approach is, however, cumbersome.
Far simpler is to discretize the field, perform all manipulations with the discrete coherent states, and revert back to the continuum for the final result.
This is the approach we adopt in the main text.

%%%%%%%%%%%%%%%%%%%%%%%%%%%%%%%
\section{The Wigner Distribution}
\label{app:wigner}
%%%%%%%%%%%%%%%%%%%%%%%%%%%%%%%

In the main text we primarily made use of the Glauber-Sudarshan quasi-probability distribution, $P(\alpha)$.
This representation has a number of uses, but when it comes to performing calculations is most useful when evaluating normally ordered quantities as shown in Eq.~\eqref{eq:opticalequivalence}.
In this appendix we briefly outline how to generalize the notion of quasi-probability distributions to different operator orderings and use this to introduce another convenient representation, the Wigner distribution.

Due to the noncommutativity of operators in a quantum theory, expectation values differ between operator ordering schemes.
As shown in Refs.~\cite{Cahill:1969it, Cahill:1969iq}, for an $s$-operator ordering we can associate the density operator with a corresponding characteristic function $\chi^{(s)}(\xi)$ specified by $\xi \in \mathbb{C}$.
In detail,
\be
\hat{\rho} = \int \frac{d\xi}{\pi}\, [\hat{D}^{(-s)} (\hat{a} , \xi)]^{\dag} \chi^{(s)}(\xi),
\ee
which is written in terms of the $(s)$-ordered displacement operator,
\be
\hat{D}^{(s)}(\hat{a}, \alpha) = e^{s |\alpha|^2 /2} \hat{D}(\hat{a}, \alpha) = e^{s |\alpha|^2 /2} e^{\alpha \hat{a}^{\dag} - \alpha^{*} \hat{a}}.
\label{eq:displacement-discrete}
\ee
In this prescription, $s = +1$ corresponds to normal ordering, $s=0$ to symmetric ordering, and $s= -1$ to antinormal ordering.
(An example of a symmetrically ordered operator is provided in Eq.~\eqref{eq:sym-example}.)
The displacement operator itself creates a coherent state with amplitude $\alpha$ from the vacuum, in particular
\be
\hat{D}(\hat{a}, \alpha) \vt 0 \ra = \vt \alpha \ra.
\label{eq:Dvac-discrete}
\ee

The characteristic function is related to a quasi-probability distribution $F^{(s)}(\alpha)$ by Fourier transform,\footnote{We emphasize once more that the Fourier transform convention we use here differs from that in the main text and App.~\ref{app:Quantum-CLT}.
Here our motivation, following Refs.~\cite{Cahill:1969it, Cahill:1969iq}, is to define a transform that maximizes the similarity with the displacement operator.
Having said that, we slightly deviate from the conventions of that work by including $1/\pi^2$ in Eq.~\eqref{eq:Fs-QPD} (rather than $1/\pi$) to ensure that $\int d\alpha\, F^{(s)}(\alpha) = 1$, as is done for $P(\alpha)$.}
\be
F^{(s)}(\alpha) = \int \frac{d\xi}{\pi^2} \, \chi^{(s)} (\xi) D(\xi, \alpha),
\label{eq:Fs-QPD}
\ee
where the displacement operator with two c-number arguments is given by $D(\xi,\alpha) = e^{\alpha \xi^*-\alpha^* \xi}$.
To help interpret these results, using subscripts $r$ and $i$ for the real and imaginary parts of $\alpha,\xi \in \mathbb{C}$, we have $\alpha \xi^* - \alpha^* \xi = 2i\alpha_i \xi_r - 2i \alpha_r \xi_i$, so that we are taking $(-\xi_i,\xi_r)$ as the conjugate variables to $(\alpha_r,\alpha_i)$.
Further, we can use this notation to represent the Dirac-$\delta$ as,
\be
\delta(\alpha) = \int \frac{d\xi}{\pi^2}\, e^{\alpha \xi^* - \xi^* \alpha} = \int \frac{d\xi}{\pi^2}\,D(\xi,\alpha).
\ee
Inverting Eq.~\eqref{eq:Fs-QPD}, we can write the density operator in terms of the quasi-probability distribution as
\be
\hat{\rho} = \int \frac{d\xi d\alpha}{\pi}\,\hat{D}^{(-s)} (\hat{a}, \xi) {D}(\xi, \alpha) F^{(s)}(\alpha). 
\ee

From here three different quasi-probability distributions can be determined for the different choices of $s$.
If we take $s=+1$, corresponding to normal ordering, the characteristic function is given by
\be
\chi^{(+)} (\xi) = \operatorname{Tr}\left[\hat{\rho}\, \hat{D}^{(+)}(\hat{a},\xi) \right]\!.
\ee
This is the characteristic function corresponding to the Glauber-Sudarshan representation, i.e. $P(\alpha) = F^{(+)}(\alpha)$.
The other convenient distribution we wish to introduce is that named after Wigner~\cite{Wigner:1932eb}, which arises for symmetric ordering.
The characteristic function is
\be
\chi^{(0)} (\xi) = \operatorname{Tr}\left[\hat{\rho}\, \hat{D}^{(0)}(\hat{a},\xi) \right]\!,
\label{eq:chi-Wigner}
\ee
and the associated quasi-probability distribution is denoted $W(\alpha) = F^{(0)}(\alpha)$.

Let us briefly collect several useful properties of $W(\alpha)$.
Firstly, when $P(\alpha)$ takes the Gaussian form, the Wigner distribution also takes the form of a Gaussian, albeit with a different variance.
This follows from substituting Eq.~\eqref{eq:GaussianRho} into first Eq.~\eqref{eq:chi-Wigner} and then Eq.~\eqref{eq:Fs-QPD}, from which we obtain
\be
W(\alpha) = \frac{1}{\pi\left(N+\tfrac{1}{2}\right)}e^{-{|\alpha|^2 }/{\left(N + \tfrac{1}{2}\right)}}.
\label{eq:WignerP}
\ee
Here and in general we normalize the Wigner distribution similarly to $P(\alpha)$, in particular $\int d\alpha\,W(\alpha)=1$.
The expression can be generalized to multiple modes in analogy to Eq.~\eqref{eq:Gauss-multimode}.

Several differences between $P(\alpha)$ and $W(\alpha)$ should be noted.
The Wigner distribution is always a continuous function of $\alpha$, whereas $P(\alpha)$ can be highly singular, as demonstrated in Eq.~(\ref{eq:numberP}).
This allows us to identify a closed-form $W(\alpha)$ for any state DM could be associated with.
However, the criteria of negativity being in one-to-one correspondence with quantum states holds only for $P(\alpha)$.

Just as $P(\alpha)$ is convenient for the calculation of normally ordered expectation values, so too $W(\alpha)$ is optimal for symmetrically ordered expressions.
In particular (cf. Eq.~\eqref{eq:opticalequivalence}),
\be
\la \{ \hat{A} ( \hat{a}, \hat{a}^{\dag} ) \}_{S} \ra 
= \int d \alpha\, W(\alpha )  A(\alpha),
\label{eq:sym_operator}
\ee
where $\{ \hat{a}^{m} (\hat{a}^{\dag})^{n} \}_{S}$ represents the average of $(m+n)! / (m! n!) $ operator orderings.
For example,
\be
\{\hat{a}^2 \hat{a}^{\dag} \}_S = 
\frac{1}{3} \left( \hat{a}^2 \hat{a}^{\dag}
+ \hat{a} \hat{a}^{\dag} \hat{a}
+ \hat{a}^{\dag} \hat{a}^2 \right)\!.
\label{eq:sym-example}
\ee
This result can be used to simplify many calculations; for instance, it is used to compute Eq.~\eqref{eq:hatphi2n} in the main text.

In the limit where $N \gg 1$ and we can approximate $[a,a^\dag] \simeq 0$ the operator ordering becomes irrelevant and we can equally well compute expectation values using $P(\alpha)$ through Eq.~\eqref{eq:opticalequivalence} or $W(\alpha)$ with Eq.~\eqref{eq:sym_operator}.
As $W(\alpha)$ is continuous it can be easier to work with.
This does not, however, allow us to circumvent the requirement of $P(\alpha) \geq 0$ to treat $\hat{\phi}$ as a classical wave in this limit.
For example, the Wigner distribution of the $N$ particle Fock state is,
\be
W(\alpha) = \frac{2}{\pi} (-1)^Ne^{-2|\alpha|^2} L_N(4|\alpha|^2),
\ee
with $L_N$ the Laguerre polynomials.
Although far less singular than the equivalent Glauber-Sudarshan distribution in Eq.~\eqref{eq:numberP}, for $N > 0$ this distribution is both positive and negative across its domain, excluding any probability based interpretation.

%%%%%%%%%%%%%%%%%%%%%%%%%%%%%%%
\section{Zero-Point Divergence}
\label{app:zeropoint}
%%%%%%%%%%%%%%%%%%%%%%%%%%%%%%%

As observed in the main text, the quantum approach to computing DM observables inevitably leads to the appearance of divergences.
The goal of this appendix is the briefly review how these can be regulated.

The divergence was first observed in Eq.~\eqref{eq:hatphi2n} and is present in the lowest non-trivial mode,
\be
\la \hat{\phi}^2\ra = \int \frac{d^3\bk}{(2\pi)^3} \frac{N_{\bk}}{\omega_{\bk}}
+ \int \frac{d^3\bk}{(2\pi)^3} \frac{1}{2\omega_{\bk}}.
\label{eq:zeropoint}
\ee
As from Eq.~\eqref{eq:Nk} $N_{\bk} \propto p(\bk)$, the first term is regulated in the UV.
The second term is not and is quadratically divergent, $\int^{\Lambda} d^3\bk/\omega_{\bk} \sim \Lambda^2$.
The divergence is not an unusual one in field theory and for an extended discussion of how to treat such divergences carefully, see e.g. Ref.~\cite{Jaffe:2005vp}.
In particular, this is similar to the usual zero-point divergence that appears in the Hamiltonian.
Note that the Hamiltonian can be written as,
\be
\hat{H} = \frac{1}{2}\int d^3\bx \left[ (\partial_t \hat{\phi})^2 + (\nabla \hat{\phi})^2 + m^2 \hat{\phi}^2 \right]\!.
\label{eq:zeropoint-H}
\ee
Substituting in the mode expansion for the fields, we find a similar divergence to that in Eq.~\eqref{eq:zeropoint}, although now with a quartic divergence, $\int^{\Lambda} d^3\bk\,\omega_{\bk} \sim \Lambda^4$.

As a first step to regulating the divergence we couple $\hat{\phi}^2$ to a classical current $J$, through ${\cal L} \supset \hat{\phi}^2 J$.
We can view $J$ as a proxy for measuring $\hat{\phi}^2$: it is an external field whose value depends on the scalar operator.
Even if we defined $\hat{\phi}^2$ to be normal ordered in order to remove the zero-point divergence, the problem is not avoided.
This is because an identical divergence appears at one loop from the new coupling $\hat{\phi}^2 J$: closing the scalar loop we obtain a diagram that is quadratically divergent in the UV and matches Eq.~\eqref{eq:zeropoint}.
Although this is a standard one-loop divergence and we can remove it by adding a counterterm $\Delta\cdot J$ to our Lagrangian.
As usual, we choose $\Delta$ to absorb the UV divergence, but even then we remain free to move finite contributions in and out of $\Delta$, which simply represents a choice of renormalization scheme.
A similar procedure can be used to regulate the Hamiltonian: we add a counterterm $\rho_{\Delta}$ to the Lagrangian, and now moving finite contributions between $\rho_{\Delta}$ and $\hat{H}$ is equivalent to choosing the zero scale of energy, and in non-gravitating theories has no physical impact.

The above procedure is sufficient to resolve all divergences the calculations in this work generate.
Focusing on $\hat{\phi}^2$, we define a partition function as
\be
Z[J] = \int [{\cal D} \phi]\, e^{i \int d^4x\, [{\cal L}_0 + (\phi^2+\Delta)J]}.
\ee
From here, we can compute the correlators of the scalar field through,
\bea
\la \hat{\phi}^{2n} \ra 
= &\left.\frac{(-i)^n}{Z[0]} \frac{\delta^n Z[J]}{\delta J^n} \right|_{J=0} \\
=\, &\frac{1}{Z[0]} \int [{\cal D} \phi]\, (\phi^2+\Delta)^n e^{i \int d^4x\, {\cal L}_0}.
\label{eq:partition-phi2n}
\eea
Importantly, this expression reveals that a choice of $\Delta$ that removes the UV divergence in $\la \phi^2 \ra$ is sufficient to remove all the divergences that appear in  Eq.~\eqref{eq:hatphi2n}, as they all appear in an identical form to the final line in Eq.~\eqref{eq:partition-phi2n}.
We thereby justify that this one regulator is sufficient and that it has no impact on the general form of the correlators.

Note that when writing $\la \hat{\phi}^{2n} \ra$ we left the specific state the system is in ambiguous.
Formally the state enters through the boundary conditions for the scalar field configurations included as part of the definition of $Z[J]$.
Commonly in field theory scattering calculations we choose this external state to be the vacuum.
In the present work, however, it is more interesting to evaluate the state between $\la \alpha \vt $ and $\vt \alpha \ra$, as after doing so we can integrate the result over $\alpha$ with weighting $P(\alpha)$ to obtain the result for a general density matrix, as justified by Eq.~\eqref{eq:GSP}.
Although technically important for the definition of $Z[J]$, if our focus is purely on regulating the divergence, the specific state is irrelevant.
In particular, if we compute $\la X \vt \hat{\phi}^2 \vt X \ra$ for any normalized state $\vt X \ra$, the divergence appearing in Eq.~\eqref{eq:zeropoint} remains identical---this is true even for the divergence appearing at one-loop, as that loop has no external $\hat{\phi}$ fields.
Accordingly, we can also fix $\Delta$ for one state and be confident that the system remains regulated even for a different density matrix.

%%%%%%%%%%%%%%%%%%%%%%%%%%%%%%%
\section{The Complex Analytic Signal and Conventions for the PSD}
\label{app:PSDconv}
%%%%%%%%%%%%%%%%%%%%%%%%%%%%%%%

There are three common conventions for the PSD: two sided, one sided, and that of the complex analytic signal.
Within this appendix we denote these as $S_2$, $S_1$, and $S_z$, respectively.
Throughout the main text we used only the one-sided PSD and in the present appendix we review the differences between these definitions and the impact on the definitions of the coherence time and volume.

The PSD captures how the power -- in the sense of the value of $\phi^2$, see Eq.~\eqref{eq:plancherel} -- is distributed over frequency.
Following the Wiener–Khinchin theorem, the PSD is the Fourier transform of autocorrelation function.
Therefore the most direct definition of the PSD is,
\be
S_2 (\omega) = \int_{-\infty}^{\infty} d\tau\,   \Gamma(\tau) e^{i \omega \tau}.
\label{eq:two_sided_PSD}
\ee
That $\Gamma(\tau)$ is even implies that we can replace $e^{i \omega \tau} \to \cos(\omega \tau)$ in the integrand and therefore as $\Gamma(\tau) \in \mathbb{R}$, we have $S_2 (\omega) \in \mathbb{R}$.
More importantly for the present discussion, the evenness of $\Gamma(\tau)$ implies that $S_2(\omega)$ is an even function of frequency, implying there is no unique information carried by the negative frequencies.

The presence of both positive and negative frequencies is a simple consequence of the Fourier transform being a decomposition in a basis of complex exponentials, $e^{i\omega \tau}$; the presence of both ensures the inverse transform produces a real function $\Gamma(\tau)$.
Nevertheless, when associating frequencies with energies in the quantum theory, it is convenient to define a PSD where we can work without the negative contributions.
The simplest approach is to define the PSD that appears in Eq.~\eqref{eq:two_sided_PSD} as two sided and then define the one-sided PSD through
\be
\frac{1}{2} S_1(\omega) = \int_{-\infty}^{\infty} d\tau\,\Gamma(\tau) e^{i\omega \tau}.
\label{eq:S1def}
\ee
Comparing to Eq.~\eqref{eq:PSD-def} we confirm that the PSD used in the main text is one sided, whereas from Eq.~\eqref{eq:two_sided_PSD}, we see the two definitions are related by $S_1(\omega) = 2 S_2(\omega)$.
Importantly, $S_1(\omega)$ still has support at negative frequencies.
The only difference is that one can obtain correct expressions for the total power when integrating over only the ``physical'' frequencies $\omega \in [0,\infty)$ (as exploited in Eq.~\eqref{eq:plancherel}).
When working with $S_2(\omega)$ one must integrate over both positive and negative frequencies to recover the entire power in the field.

The final PSD we discuss is that associated with the complex analytic signal, $S_z$, and is defined such that for $\omega > 0$, $S_z(\omega) = S_2(\omega)$, whereas for $\omega < 0$, $S_z(\omega)=0$.
Our discussion partially follows Ref.~\cite{Mandel:1995seg} and we refer there for further details.
To ensure the PSD has no support at negative values, the associated autocorrelation function, denoted $\Gamma_z(\tau)$, must be complex.
More fundamentally, let us focus on the case where the quantum field of interest is well approximated by its classical field analog, $\phi(t)$, given in Eq.~\eqref{eq:phicnumber}.
(We discard the positional dependence of the field for most of the discussion to simplify the notation.)
If we denote the Fourier transform of the field as $\tilde{\phi}(\omega)$, then the associated complex analytic signal is defined by
\be
\phi_z(t) = \int_0^{\infty} \frac{d\omega}{2\pi}\, \tilde{\phi}(\omega) e^{-i \omega t},
\label{eq:phiz-def}
\ee
such that we simply remove the negative frequencies when performing the inverse Fourier transform.
(As a simple example, if $\phi(t) = \cos(mt)$ then $\phi_z(t) = \tfrac{1}{2} e^{-imt}$.)
Note that the original function can be recovered directly through $\phi(t) = 2\operatorname{Re}\phi_z(t)$.
Although we focus on the classical field approximation, there is a close connection between the complex analytic signal and the decomposition of the field operator into positive and negative frequency modes as performed in Eq.~\eqref{eq:phi-plusminus}.
In particular, if we apply the definition in Eq.~\eqref{eq:phiz-def} to the discrete mode decomposition of the field in Eq.~\eqref{eq:phi-finiteV}, we find $\hat{\phi}_z(t) = \hat{\phi}^+(t)$.

Continuing, we define the associated autocorrelation function as,
\be
\la \Gamma_z(\tau) \ra = \la \phi_z^*(t) \phi_z(t+\tau) \ra.
\label{eq:Gamz-def}
\ee
Assuming strong stationarity we can compute
\be
\la \Gamma_z (\tau)  \ra = \frac{\bar{n}}{2} \int_0^{\infty} d \omega\, \frac{p(\omega)}{\omega} e^{-i \omega \tau}.
\label{eq:Gz-explicit}
\ee
Using $\la \Gamma(\tau) \ra = 2\operatorname{Re} \la \Gamma_z(\tau) \ra$ we recover Eq.~\eqref{eq:Gamma-tau}; note also $\la \Gamma(0) \ra = 2\la \Gamma_z(0) \ra$.
These results should be compared to Eqs.~\eqref{eq:g1pk} and \eqref{eq:g1-2-Gam}.
The above discussion reveals that $g^{(1)}(\tau) = \la\Gamma_z(\tau) \ra/\la \Gamma_z(0) \ra$, a result which could also be seen from the fact that an equivalent definition for $\Gamma_z(\tau)$ in Eq.~\eqref{eq:Gamz-def} can be obtained from the quantum fields, $\la \Gamma_z (\tau) \ra = \la \hat{\phi}^-(t) \hat{\phi}^+(t+\tau) \ra$, cf. Eq.~\eqref{eq:def-g1g2}.
Finally, the PSD can be determined in two ways,\footnote{The second relation also holds with $S_z \to S_2$ and $\tilde{\phi}_z \to \tilde{\phi}$, although it is often more convenient to compute the PSD as the Fourier transform of $\Gamma(\tau)$.}
\bea
\la S_z(\omega) \ra &= \int_{-\infty}^{\infty}d\tau\, \la\Gamma_z(\tau) \ra  e^{i \omega \tau}, \\
\la \tilde{\phi}^*_z(\omega) \tilde{\phi}_z(\omega') \ra &= 2\pi \la S_z(\omega) \ra \delta(\omega-\omega').
\eea
As $\tilde{\phi}_z(\omega)$ vanishes for $\omega < 0$, the second expression demonstrates that $S_z$ only has support for positive frequencies.
From Eq.~\eqref{eq:Gz-explicit} we can compute,
\be
\la S_z(\omega) \ra = \frac{\pi \bar{n}p(\omega)}{\omega},
\ee
where now $p(\omega)$ should only be considered to have positive support.
We can connect with Eq.~\eqref{eq:PSD} as $S_2(\omega) = S_z(\omega) + S_z(-\omega)$ and again $S_1(\omega) = 2 S_2(\omega)$.
All three PSDs are related through
\be
\int_{-\infty}^{\infty} d\omega\, S_2(\omega) 
= \int_0^{\infty} d\omega\, S_1(\omega) 
= 2 \int_0^{\infty} d\omega\, S_z(\omega). 
\label{eq:App_PSDrelation}
\ee

Having introduced the various PSD conventions we next turn to a discussion of their relation to the definition of the coherence time and volume.
A common definition of the coherence time in quantum optics is~\cite{Mandel:1995seg}
\be
\tau_c = \int_{-\infty}^{\infty} d\tau\,\left|\frac{\left\langle\Gamma_z(\tau)\right\rangle}{\left\langle\Gamma_z(0)\right\rangle}\right|^2\!.
\label{eq:def-tc-z}
\ee
The advantage of this definition is that it completely removes the oscillatory behavior of the autocorrelation function and associates $\tau_c$ with the decay of its amplitude.
Nevertheless, one can show that if the system obeys strong stationarity, we have
\be
\int_{-\infty}^{\infty } d\tau\,  | \la \Gamma_{z} (\tau) \ra|^2 
= \frac{1}{2} \int_{-\infty}^{\infty} d\tau\, \la \Gamma(\tau) \ra^2.
\ee
which together with $\la \Gamma(0) \ra = 2\la \Gamma_z(0) \ra$ establishes that Eq.~\eqref{eq:def-tc-z} is equivalent to the definition of the coherence time using in the main text, Eq.~\eqref{eq:tauc-def}.
Given the connection between $g^{(1)}$ and $\Gamma_z$ discussed above, we further justify the alternative definition given in Eq.~\eqref{eq:tcVc-g1}.

Restoring the positional dependence to the fields, the natural definition of the coherence volume is apparent and indeed has already been stated in Eq.~\eqref{eq:tcVc-g1}.
We can further confirm why the two definitions given in Eq.~\eqref{eq:Vc-def} do not exactly match.
The problem lies with the oscillatory contributions that remain when working with $\Gamma(\bd)$ rather than $\Gamma_z(\bd)$.
In particular, using Eq.~\eqref{eq:Gamma-tau-d} we have
\bea
2 &\int d^3\bd\,\left[ \frac{\la \Gamma(\bd) \ra}{\la \Gamma(\mathbf{0}) \ra} \right]^2  \\
& = \frac{(2 \pi)^3}{\la 1/ \omega \ra^2} 
\int_{-\infty}^{\infty} \frac{d^3 \bk}{\omega_{\bk}^2} \left[ p(\bk)^2 + p(\bk) p(-\bk) \right]\!.
\eea
In general $p(\bk)$ is not even, so the terms cannot be combined and so it is clear that the second expression given for $V_c$ in Eq.~\eqref{eq:Vc-def} slightly differs from that computed with the complex analytic signal.
Note the SHM of Eq.~\eqref{eq:SHM} is explicitly not even in $\bk$, although direct computation reveals the two definitions give similar values nonetheless.

As a final note, we point out that the various PSD conventions have not always been used consistently in the literature.
For example, Ref.~\cite{Foster:2017hbq} neglects negative frequencies, although does not include the factor of two in Eq.~\eqref{eq:S1def} as appropriate for a one-sided PSD.
Therefore, formally, the PSD as used in that reference is a factor of two too small, which would propagate to the definition of the axion likelihood provided in that work, and implemented for the experimental analyses used by ABRACADABRA~\cite{Ouellet:2018beu,Ouellet:2019tlz,Salemi:2021gck}.
Nonetheless, all experimental sensitivities are computed using likelihood ratios, in which the neglected factor exactly cancels, leaving the results unaltered.

%%%%%%%%%%%%%%%%%%%%%%%%%%%%%%%
\section{Full Calculation Through the Wave-Particle Transition}
\label{app:WP-Full}
%%%%%%%%%%%%%%%%%%%%%%%%%%%%%%%

In this appendix we provide a detailed calculation for the mean and variance of the energy of a non-relativistic scalar field within an arbitrary volume $V$, assuming a Gaussian $P(\alpha)$.
The results of this calculation were discussed in Sec.~\ref{sec:boundary} and we refer there for additional details as well as a discussion of how these results extend to higher moments.
Beyond those specifics, the calculation in this appendix also provides an explicit example of a full quantum calculation where results can be extracted independent of the assumed $N$.
It further demonstrates explicitly how the coherence properties of the field emerge naturally in the calculation, rather than being heuristic properties one associated with the fields evolution.

To begin with, for a non-relativistic field the density operator is given by
\bea
\hat{\rho}(t,\bx) =\,& \frac{1}{2} \left[ (\partial_t \hat{\phi})^2 + (\nabla \hat{\phi})^2 + m^2 \hat{\phi}^2 \right] \\
=\,&\sum_{\bk,\bq} \frac{m}{\cal V}\, \hat{a}_{\bk}^\dag \hat{a}_{\bq} e^{i (k-q) \cdot x},
\label{eq:rhohat-NR}
\eea
where we have removed the zero-point divergence as discussed in App.~\ref{app:zeropoint} (indeed $\hat{H} = \int d^3\bx\,\hat{\rho}$).
In order to avoid confusion with the physical volume we are studying, $V$, within this appendix only we denote the volume that is used to discretize the field as ${\cal V}$.
In terms of this, the energy operator is simply the density integrated over the volume of interest,
\be
\hat{M}(t) = \int_V d^3\bx\,\hat{\rho}(t,\bx).
\ee
Let us make several comments already.
Firstly, as we are dealing with a non-relativistic field, the energy in the volume is equivalent to the enclosed mass and so we have defined the operator with the symbol $\hat{M}$.
Building on this, we emphasize that while in the non-relativistic limit all modes have energy $\omega_{\bk} = m$ up to ${\cal O}(v^2)$ corrections -- as assumed in Sec.~\ref{sec:boundary} -- we retain the energy differences in the phase factor that enters Eq.~\eqref{eq:rhohat-NR}; as we see explicitly below, the phases are where the coherence volume emerges from.
Further, in general the operators explicitly depend on time, however, as our focus is on the spatial fluctuations of the energy we set $t=0$ and remove the dependency moving forward.
How fluctuations in time should be considered was discussed in Sec.~\ref{sec:boundary}.
Lastly, for computational convenience we take $V = L^3$ to be a cubic region and we leave the fact that the region being integrated over is $V$ implicit moving forward.

Having established our conventions, the mean value follows directly, although being explicit
\begin{align}
\la \hat{M} \ra 
=\,& \int d^3\bx\,\la \hat{\rho}(\bx) \ra 
= \int d^3\bx\,\sum_{\bk,\bq} \frac{m}{\cal V}\, \la\hat{a}_{\bk}^\dag \hat{a}_{\bq} \ra e^{i (\bq-\bk) \cdot \bx} \nonumber \\
=\,&\int d^3\bx\,\sum_{\bk} \frac{m}{\cal V}\, N_{\bk}
=\,m \bar{n} V.
\end{align}
Accounting for the energy of all modes being $m$, this then matches the mean in Eq.~\eqref{eq:musig-generalV}.

The variance can be determined from $\la \Delta \hat{M}^2 \ra = \la \hat{M}^2 \ra - \la \hat{M} \ra^2$, so that what remains is to compute the second moment.
Proceeding as above,
\bea
\la \hat{M}^2\ra = \int d^3\bx\,d^3\bx' 
&\sum_{\bk,\bq,\bk',\bq'} \frac{m^2}{{\cal V}^2} \big\la\hat{a}_{\bk}^\dag \hat{a}_{\bq} \hat{a}_{\bk'}^\dag \hat{a}_{\bq'} \big\ra \\
\times\, &e^{i [(\bq-\bk) \cdot \bx + (\bq'-\bk') \cdot \bx']}.
\label{eq:fc-wp-msq}
\eea
Consider the expectation value,
\bea
\big\la\hat{a}_{\bk}^\dag \hat{a}_{\bq} \hat{a}_{\bk'}^\dag \hat{a}_{\bq'} \big\ra
=\,& \big\la\hat{a}_{\bk}^\dag \hat{a}_{\bk'}^\dag \hat{a}_{\bq} \hat{a}_{\bq'} \big\ra + \big\la\hat{a}_{\bk}^\dag \hat{a}_{\bq'} \big\ra \delta_{\bk',\bq} \\
=\,&\big\la \alpha_{\bk}^* \alpha_{\bk'}^* \alpha_{\bq} \alpha_{\bq'} \big\ra + \big\la\alpha_{\bk}^* \alpha_{\bq'} \big\ra \delta_{\bk',\bq}.
\eea
To evaluate this, we observe that as the $\alpha$ values are normally distributed, we can apply Wick's theorem to the first expression, giving
\begin{align}
\big\la \alpha_{\bk}^* \alpha_{\bk'}^* \alpha_{\bq} \alpha_{\bq'} \big\ra
=\,&\big\la \alpha^*_{\bk} \alpha^*_{\bk'} \big\ra \big\la \alpha_{\bq} \alpha_{\bq'} \big\ra
+  \big\la \alpha^*_{\bk} \alpha_{\bq} \big\ra \big\la \alpha^*_{\bk'} \alpha_{\bq'} \big\ra \nonumber \\
+\,&\big\la \alpha^*_{\bk} \alpha_{\bq'} \big\ra \big\la \alpha_{\bq} \alpha^*_{\bk'} \big\ra.
\end{align}
The first term vanishes, leaving
\bea
\big\la\hat{a}_{\bk}^\dag \hat{a}_{\bq} \hat{a}_{\bk'}^\dag \hat{a}_{\bq'} \big\ra
=\,& N_{\bk} N_{\bk'} \delta_{\bk,\bq} \delta_{\bk',\bq'} \\
+\,& N_{\bk} (N_{\bk'}+1) \delta_{\bk,\bq'} \delta_{\bk',\bq}.
\label{eq:4a-exp}
\eea
Substituting the first line of this result back into Eq.~\eqref{eq:fc-wp-msq}, we obtain $\la \hat{M} \ra^2$.
Therefore, the second line of Eq.~\eqref{eq:4a-exp} specifies the variance, which after moving to the continuous representation for the modes becomes
\bea
\la \Delta \hat{M}^2 \ra =\, &\frac{m^2\bar{n}}{(2\pi)^3} \int d^3\bx\,d^3\bx' \int d^3\bk\, d^3\bq \\
\times\, &p(\bk)\, [(2\pi)^3 \bar{n}\, p(\bq)+1] e^{i (\bq-\bk) \cdot (\bx-\bx')}.
\label{eq:fc-wp-midpoint}
\eea
There are two distinct integrals over momenta associated with the two terms in square brackets.
Starting with the first and introducing the shorthand $\bd=\bx-\bx'$, we have
\bea
&\bar{n} \int d^3\bk\, d^3\bq\, p(\bk)\, p(\bq) e^{i (\bq-\bk) \cdot \bd} \\
= &\bar{n} \left[ \int d^3\bk\, p(\bk) e^{-i \bk \cdot \bd} \right] \left[ \int d^3\bq\, p(\bq) e^{i \bq \cdot \bd} \right] \\
= &\bar{n} \left|\frac{\left\langle\Gamma_z(\bd)\right\rangle}{\left\langle\Gamma_z(\mathbf{0})\right\rangle}\right|^2 
= \bar{n} \big|g^{(1)}(0,\bd)\big|^2,
\eea
where we observed that the integrals over momentum could be rewritten in terms of the complex analytic signal discussed in App.~\ref{app:PSDconv} and then exploited its connection to $g^{(1)}$ introduced in Sec.~\ref{sec:g1g2}.
The second integral from Eq.~\eqref{eq:fc-wp-midpoint} is more straightforward,
\bea
\frac{1}{(2\pi)^3}\int d^3\bk\, d^3\bq\, p(\bk) e^{i (\bq-\bk) \cdot \bd} = \delta(\bd),
\eea
Returning to the full expression in Eq.~\eqref{eq:fc-wp-midpoint}, as the integrand depends only on the difference of the positions through $\bd$, we can change variables to reduce this to,
\be
\la \Delta \hat{M}^2 \ra = m^2\bar{n}V\, [ \bar{n} \theta(V) + 1 ],
\ee
with
\bea
\theta(V) =\, &\int d^3\bd\, T(\bd)\, \big|g^{(1)}(0,\bd)\big|^2, \\
T(\bd) \equiv\, &V^{-1}(L-|d_x|) (L-|d_y|) (L-|d_z|).
\eea
where the integral over $\bd$ is also performed over the volume of interest $V$.
Accounting for the energy of the modes, this justifies the variance stated in Eq.~\eqref{eq:musig-generalV}.
Note a similar expression to $\theta(V)$ appears when studying fluctuations of the field over arbitrary times, as compared to $\tau_c$, as considered in Ref.~\cite{MANDEL1963181} (see also Ref.~\cite{Mandel:1995seg} which adopted the $\theta$ notation we follow).

The asymptotic behavior of $\theta(V)$ is independent of $g^{(1)}(0,\bd)$.
For $V \ll V_c$, $\bd$ is restricted to a region where $|g^{(1)}(0,\bd)| \simeq 1$ (cf. Fig.~\ref{fig:Gam-d-tau}).
Therefore,
\be
\lim_{V \ll V_c} \theta(V) \simeq \int_V d^3\bd\, T(\bd) = V.
\ee
Secondly, for $V \gg V_c$, $|g^{(1)}(0,\bd)|^2$ decays rapidly when $\bd$ is outside the coherence volume, such that $T(\bd) \simeq 1$ and we now have
\be
\lim_{V \gg V_c} \theta(V) \simeq \int_V d^3\bd\, \big|g^{(1)}(0,\bd)\big|^2 = V_c,
\ee
as this is exactly the definition of the coherence volume in Eq.~\eqref{eq:tcVc-g1}.
This confirms that the asymptotics seen in Fig.~\ref{fig:Theta-V} are general, whereas the transition between them depends on the exact form of $g^{(1)}$.
Using the SHM as a specific example, the analytic form for $\big|g^{(1)}(0,\bd)\big|^2$ was given in Eq.~\eqref{eq:g12-SHM} and from this we can evaluate
\be
\frac{\theta_{\DM}(V)}{V_c} = \frac{\Big[ \pi \nu^{1/3} \operatorname{Erf} \!\big( \sqrt{\pi} \nu^{1/3} \big)\! - 1 + e^{-\pi \nu^{2/3}} \Big]^3}{\pi^3 \nu},
\ee
with $\nu=V/V_c$.
This is exactly the form that was shown in Fig.~\ref{fig:Theta-V} and in particular we have $\theta_{\DM}(V_c) \simeq 0.319 V_c$, consistent with Eq.~\eqref{eq:thetaV-limits}.

%%%%%%%%%%%%%%%%%%%%%%%%%%%%%%%
\section{An Example of the Evolution of $P(\alpha)$}
\label{app:Pa-Evo}
%%%%%%%%%%%%%%%%%%%%%%%%%%%%%%%

As emphasized in the main text, a full justification of the assumed Gaussian $P(\alpha)$ requires understanding how the density matrix of DM evolved.
Here we review an example of how one can evolve $P(\alpha)$ in a highly idealized scenario, one where the DM is coupled to a thermal reservoir.
We emphasize that in no sense is this intended to be a model for how the $P(\alpha)$ of DM evolved in our Universe and the final result is unsurprising: $P(\alpha)$ evolves to a Gaussian as the system is thermalizing and a thermal distribution is equivalent to a Gaussian $P(\alpha)$.
Our review follows closely Refs.~\cite{Carmichael:1999,Cao:2022kjn} and we refer to those works for further details.
(See also Ref.~\cite{Allali:2020ttz, Allali:2020shm, Allali:2021puy} for a similar study.)

Formally, the evolution of the system is determined by a master equation for the density matrix.
Restricting our attention to a single mode of frequency $m$, when the system is coupled to a thermal reservoir the master equation takes on the Lindblad form,
\begin{align}
\dot{\hat{\rho}} \simeq& - i m [\hat{a}^{\dag} \hat{a},~\hat{\rho}] + \frac{\gamma}{2} (\bar{n}+1) (2 \hat{a} \hat{\rho} \hat{a}^{\dag} - \hat{a}^{\dag} \hat{a} \hat{\rho} - \hat{\rho} \hat{a}^{\dag} \hat{a} ) \nonumber\\
& + \frac{\gamma}{2} \bar{n} (2\hat{a}^{\dag} \hat{\rho} \hat{a}^{\dag} - \hat{a} \hat{a}^{\dag} \hat{\rho} - \hat{\rho} \hat{a} \hat{a}^{\dag}).
\label{eq:mastereq}
\end{align}
This result, which is effectively the equation of a damped harmonic oscillator, is stated in the Schrodinger picture.
The properties of the reservoir are encoded in the average occupation at the mode of interest, $\bar{n} = [e^{m/T}-1]^{-1}$, and the dissipation coefficient, $\gamma$, is determined by the interaction between the DM and the reservoir.
It is through these interactions that the DM field inherits the noise from the environment that drives its evolution.

The goal is to describe the evolution of $P(\alpha,t)$ as a function of time.
Substituting in Eq.~\eqref{eq:GSP} the master equation of Eq.~\eqref{eq:mastereq} becomes a Fokker-Planck equation,
\bea
\partial_t P(\alpha, t) \simeq&  \Big[\left(\frac{\gamma}{2} + i m \right) \partial_{\alpha} \alpha + \left(\frac{\gamma}{2} - i m  \right) \partial_{\alpha^*} \alpha^{*}  \\ 
&+ \gamma \bar{n} \partial_{\alpha} \partial_{\alpha^*} \Big]P(\alpha, t).
\label{eq:FPeq}
\eea
As an explicit example, suppose the DM mode starts in a pure coherent state, $P(\alpha, t=0) = \delta (\alpha- \alpha_{0})$.
The general solution can then be determined to be,
\bea
P(\alpha, t) = \frac{1}{\pi  \bar{n} \big(1 - e^{-\gamma t} \big)}
\exp \left[ - \frac{\big| {\alpha} - {\alpha}_{0} e^{- \gamma t/2}e^{- im t} \big|^2 }{\bar{n} \big(1 - e^{-\gamma t} \big)} \right]\!.
\label{eq:FPGaussian}
\eea
Qualitatively, the interactions between the initial coherent state and the thermal reservoir is simultaneously damping the field amplitude and injecting noise into the coherent state, leading it towards a Gaussian.
For $t \gg \gamma^{-1}$, we see that the state has evolved to a Gaussian as in Eq.~\eqref{eq:GaussianRho}, with $N=\bar{n}$.
In other words, the system has thermalized with the reservoir.

Lastly, we can use the above example as an opportunity to expand on the point that the Gaussian $P(\alpha)$ corresponds to a density matrix that is diagonal in the Fock basis, as seen in Eq.~\eqref{eq:rho-Fockbasis}.
This is far from generic.
A pure coherent state $\vt \alpha_0 \ra \la \alpha_0 \vt$ has support for arbitrary off-diagonal terms,
\be
\hat{\rho} = e^{-|\alpha_0|^2} \sum_{n,k=0}^{\infty} \frac{\alpha_0^n(\alpha_0^*)^k}{\sqrt{n!k!}} \vt n \ra \la k \vt.
\label{eq:rho-purecohere-Fock}
\ee
As another way to view this, neglecting interactions, the number operator commutes with the Hamiltonian and therefore the Fock basis can also be viewed as the energy basis: the Gaussian is diagonal in this space whilst generic states are not.
Using Eq.~\eqref{eq:FPGaussian} we can explicitly compute the evolution of both the diagonal and off-diagonal contributions to Eq.~\eqref{eq:rho-purecohere-Fock} as the system thermalizes, which can be done analytically in terms of sums of Hermite polynomials.
Doing so one finds that once $t \sim \gamma^{-1}$ the off-diagonal terms begin to be suppressed leaving only the Gaussian supported diagonal elements.
Although we have only studied this in a simple model, the diagonalization in energy will be a generic feature of the evolution of systems towards a Gaussian $P(\alpha)$.

%%%%%%%%%%%%%%%%%%%%%%%%%%%%%%%%%%%%%%%%%%%%%%%%%%
\bibliographystyle{utphys}
\bibliography{refs}
%%%%%%%%%%%%%%%%%%%%%%%%%%%%%%%%%%%%%%%%%%%%%%%%%%

\end{document}